\documentclass[5p,times]{elsarticle}
\usepackage[english]{babel}
\usepackage{hyperref}
\usepackage{comment}
\usepackage{soul}
\usepackage{color}
\usepackage{graphicx}
\usepackage{stackengine}
\usepackage{amsmath,amssymb}
\usepackage{soul}
\usepackage{tikz}
\usepackage[markup=underlined]{changes}
\definechangesauthor[color=blue]{JR}
\definechangesauthor[color=green]{WC}
\usepackage{lineno}

\begin{document}

\title{The isospin and neutron-to-proton excess dependence of short-range correlations}

\author[SRC]{Jan Ryckebusch}%
\ead{Jan.Ryckebusch@UGent.be}
\author[SRC]{Wim Cosyn}
\ead{Wim.Cosyn@UGent.be}

\author[SRC]{Sam Stevens}

\author[SRC]{Corneel Casert}
\ead{Corneel.Casert@UGent.be}

\author[SRC]{Jannes Nys}
\ead{Jannes.Nys@UGent.be}
\address[SRC]{%
 Department of Physics and Astronomy, Ghent University, Belgium
}%


\date{\today}

\begin{abstract}
We provide a systematic study of the isospin composition and neutron-to-proton $\left( \frac{N}{Z} \right)$ ratio dependence of nuclear short-range correlations (SRC) across the nuclear mass table. We use the low-order correlation operator approximation (LCA) to compute 
the SRC contribution to the single-nucleon momentum distributions for 14 different nuclei from $A=4$ to $A=208$.  Ten asymmetric nuclei are included for which the neutrons outnumber the protons by a factor of  up to 1.54.    
The computed momentum distributions are used to extract the pair composition of the SRC.
We find that there is a comprehensive picture for the isospin composition of SRC and their evolution with nucleon momentum. We also compute the non-relativistic kinetic energy of neutrons and protons and its evolution with nuclear mass $A$ and  $\frac{N}{Z}$. Confirming the conclusions from alternate studies it is shown that the minority species (protons) become increasingly more short-range correlated as the neutron-to-proton ratio  increases. We forge connections between measured nucleon-knockout quantities sensitive to SRC and single-nucleon momentum distributions. It is shown that the LCA can account for the observed trends in the data, like the fact that in neutron-rich nuclei the protons are responsible for an unexpectedly large fraction of the high-momentum components. 
\end{abstract}

\maketitle
\section{Introduction}
\label{sec:intro}
Nuclear long-range and short-range correlations (SRC) are marked by different energy-momentum scales. Long-range correlations typically occur at ``low'' (order of MeV) nuclear excitation energies and, as they extend over the size of the nucleus, relatively low nucleon momenta are involved. The majority of high-momentum nucleons, on the other hand, are born in SRC.   Inter-nucleon  SRC that correct the  independent-particle model (IPM) for nuclei for the finite size of nucleons have been introduced ever since the initiation of nuclear physics. One often refers to those correlations  as Jastrow or central correlations. SRC have been under the microscope for many years and one of the more recent insights is that the isospin-blind central correlations are not the dominant source of SRC  (see for example \cite{Schiavilla:2006xx, Piasetzky:2006ai, Alvioli:2007zz, Rios:2013zqa, Atti:2015eda, Hen:2016kwk}). Thanks to dedicated experimental and theoretical work, there are strong indications that tensor correlations play a dominant role.  Due to their isospin dependence, they are mostly affecting proton-neutron (pn) pairs.

Close to stability, indirect indications for SRC stem from quasi-free $A(e,e^{\prime}p)$ \cite{Lapika:1903} and $A(p,pp)$ data \cite{Atar:2018dhg} compared to state-of-the-art reaction-model calculations. Indeed, the ratio of measured to computed cross sections $\frac{\sigma_{exp}}{\sigma_{th}}$ is systematically of the order 0.5-0.7. In weakly bound systems the situation is less clear. The analysis of heavy-ion induced knockout reactions alludes to a very strong dependence of the spectroscopic factors on the proton-to-neutron ratio ~\cite{Gade:2008zz}. For heavy-ion induced knockout reactions there are some issues with the  reaction-model calculations and recent $A(p,pp)$ data \cite{Atar:2018dhg} do not seem to corroborate these findings with regard to the $\frac{N}{Z}$ dependence of the spectroscopic factors. In addition, long-range correlations also have a large impact on the  $\frac{\sigma_{exp}}{\sigma_{th}}$~\cite{Dickhoff:2004xx} which complicates the extraction of SRC information from quasi-free single-nucleon knockout data.    

The hallmark of SRC is two-nucleon (2N) knockout in peculiar kinematics that probes considerably higher excitation energies of the target nucleus than the aforementioned single-nucleon knockout reactions. Accordingly, 2N knockout studies provide a more direct access to SRC and have been the subject of dedicated studies with electron beams since the late 1990s~\cite{Blomqvist:1998gq, Onderwater:1998zz, Starink:2000qhh, Subedi:2008zz, Shneor:2007tu, Korover:2014dma}. Indeed, SRC-susceptible reactions  require a probe with high virtuality (to enhance the sensitivity to short-distance phenomena and suppress contributions stemming from meson-exchange processes). In addition, one needs a trigger  that selects scattering from a fast nucleon that is part of a nucleon pair with low center-of-mass momentum and a momentum configuration approaching a back-to-back situation. 

The CLAS collaboration at Jefferson Lab has studied a lot of systematic properties of SRC for the target nuclei $ ^{12}$C, $^{27}$Al, $^{56}$Fe and $^{208}$Pb.  Following the thorough analysis of $A(e,e^{\prime}p (N))$ and $A(e,e^{\prime}n)$ reactions with those nuclei one could determine the evolution with the nuclear mass number $A$~\cite{Hen:2014nza}, gain insight in its neutron-to-proton asymmetry dependence~\cite{Duer2018}, determine the quantum numbers of the short-range correlated pairs~\cite{Colle:2015ena}, and determine the center-of-mass (c.m.) properties of short-range correlated pp pairs~\cite{Cohen:2018gzh}. As the  determination of multidimensional cross sections in large phase spaces is a very tedious task,  the collaboration adopts a strategy that very much hinges on the determination of cross-section ratios for the extraction of the physical properties from the data.  

There are different strategies to compute the effect of SRC on nuclear properties for finite nuclei. For light nuclei, Monte Carlo calculations based on the Schr\"{o}dinger equation with a realistic inter-nucleon interaction, are feasible for determining the ground-state properties. Due to their enormous computational cost and the curse of dimensionality, those calculations cannot go beyond $A=40$ to date~\cite{Carlson:2014vla, Wiringa:2013ala, Lonardoni:2017egu}. This is also the highest $A$ for which cluster calculations have been reported by the Perugia group~\cite{Alvioli:2016wwp}. Here, we resort to an effective method for computing the effect of SRC that is applicable across the nuclear mass table and has the advantage that it allows for systematic studies of SRC properties with a reasonable computational cost. Central to our approach is a methodology that allows one to compute the SRC contribution to nuclear matrix elements. The method can be used for the computation of nuclear momentum distributions \cite{Vanhalst:2014cqa} and in reaction-theory studies for SRC-sensitive processes \cite{Colle:2015ena, Ryckebusch:1997gn,  VanCuyck:2016fab, Colle:2015lyl, Stevens:2017orj}. In this work we start from single-nucleon momentum distributions for a range of nuclei. As their SRC part is dominated by 2N correlations, a connection can be made to recent results of experimental studies that seek for SRC sensitive processes through the selection of 2N knockout events matching the kinematic conditions of the major SRC contributions to the single-nucleon momentum distribution.  

\begin{figure}[ht!]
    \centering
    \begin{tikzpicture}[scale=0.8,very thick]
\node (lab) at (3.8,2.5) {\textbf{\textit{(a)}}} ;
\node [red] (A) at (1.3,3) {$\vec{p}$} ;
\node [red] (B) at (2.7,3) {$\vec{p}$} ; 
\node (C) at (1.3,2.2) {$\vec{r}_1$} ; 
\node (D) at (2.7,2.2) {$\vec{r}_1^{\; \prime}$};
 \draw[cyan,fill=cyan] (0,0) -- (4,0) -- (4,2) -- (3,2) -- (3,1.5) -- (1,1.5) -- (1,2) -- (0,2) --cycle  ;
 \node (DD) at (1.9,0.4) {\large{A-1}} ;
 \node (EE) at (.5,0.4) {\large{A}} ;
 \node (FF) at (3.5,0.4) {\large{A}} ;
 \draw[red,->] (1,2) -- (1,3) ; 
 \draw[red,->] (3,3) -- (3,2) ;
    \end{tikzpicture}
\hspace{0.1\columnwidth}
%
%
    \begin{tikzpicture}[scale=0.8,very thick]
\node (lab) at (3.8,2.5) {\textbf{\textit{(b)}}} ;
\node [red] (A) at (1.3,3) {$\vec{p}$} ;
\node [red] (B) at (2.7,3) {$\vec{p}$} ; 
\node (C) at (1.3,2.2) {$\vec{r}_1$} ; 
\node (D) at (2.7,2.2) {$\vec{r}_1^{\; \prime}$};
\node (N1) at (0.1,2.2) {N$^{\phantom{\prime}}$} ;
\node (N2) at (0.1,1.7) {N$^{\prime}$} ;
 \draw [dashed] (0,0) -- (4,0);
 \draw [dashed] (0,0.5) -- (4,0.5);
 \draw [dashed] (0,1) -- (4,1);
 \draw [dashed] (0,1.5) -- (4,1.5);
 \draw [dashed] (0,2) -- (1,2);
 \draw [dashed] (3,2) -- (4,2);
 \draw[red,->] (1,2) -- (1,3) ; 
 \draw[red,->] (3,3) -- (3,2) ;
    \end{tikzpicture}
    
   \vspace{0.02\textheight}

    \begin{tikzpicture}[scale=0.8,very thick]
\node (lab) at (3.8,2.5) {\textbf{\textit{(c)}}} ;
\node [red] (A) at (1.3,3) {$\vec{p}$} ;
\node [red] (B) at (2.7,3) {$\vec{p}$} ; 
\node (C) at (1.3,2.2) {$\vec{r}_1$} ; 
\node (D) at (2.7,2.2) {$\vec{r}_1^{\; \prime}$};
\node (CD) at (2,1.85) {$\vec{r}_2$} ;
\node (N1) at (0.1,2.2) {N$^{\phantom{\prime}}$} ;
\node (N2) at (0.1,1.7) {N$^{\prime}$} ;
 \draw [dashed] (0,0) -- (4,0);
 \draw [dashed] (0,0.5) -- (4,0.5);
 \draw [dashed] (0,1) -- (4,1);
 \draw [dashed] (0,1.5) -- (4,1.5);
 \draw [dashed] (0,2) -- (1,2);
 \draw [dashed] (3,2) -- (4,2);
 \draw[red,->] (1,2) -- (1,3) ; 
 \draw[red,->] (3,3) -- (3,2) ;
 \draw[olive,dotted] (2,1.5) -- (1,2) ;
    \end{tikzpicture}
\hspace{0.1\columnwidth}
    \begin{tikzpicture}[scale=0.8,very thick]
\node (lab) at (3.8,2.5) {\textbf{\textit{(d)}}} ;
\node [red] (A) at (1.3,3) {$\vec{p}$} ;
\node [red] (B) at (2.7,3) {$\vec{p}$} ; 
\node (C) at (1.3,2.2) {$\vec{r}_1$} ; 
\node (D) at (2.7,2.2) {$\vec{r}_1^{\; \prime}$};
\node (CD) at (2,1.85) {$\vec{r}_2$} ;
\node (N1) at (0.1,2.2) {N$^{\phantom{\prime}}$} ;
\node (N2) at (0.1,1.7) {N$^{\prime}$} ;
 \draw [dashed] (0,0) -- (4,0);
 \draw [dashed] (0,0.5) -- (4,0.5);
 \draw [dashed] (0,1) -- (4,1);
 \draw [dashed] (0,1.5) -- (4,1.5);
 \draw [dashed] (0,2) -- (1,2);
 \draw [dashed] (3,2) -- (4,2);
 \draw[red,->] (1,2) -- (1,3) ; 
 \draw[red,->] (3,3) -- (3,2) ;
 \draw[olive,dotted] (2,1.5) -- (3,2) ;
    \end{tikzpicture}
    
    \vspace{0.02\textheight}
    
    \begin{tikzpicture}[scale=0.8,very thick]
\node (lab) at (3.8,2.5) {\textbf{\textit{(e)}}} ;
\node [red] (A) at (1.3,3) {$\vec{p}$} ;
\node [red] (B) at (2.7,3) {$\vec{p}$} ; 
\node (C) at (1.3,2.2) {$\vec{r}_1$} ; 
\node (D) at (2.7,2.2) {$\vec{r}_1^{\; \prime}$};
\node (CD) at (2,1.85) {$\vec{r}_2$} ;
\node (N1) at (0.1,2.2) {N$^{\phantom{\prime}}$} ;
\node (N2) at (0.1,1.7) {N$^{\prime}$} ;

 \draw [dashed] (0,0) -- (4,0);
 \draw [dashed] (0,0.5) -- (4,0.5);
 \draw [dashed] (0,1) -- (4,1);
 \draw [dashed] (0,1.5) -- (4,1.5);
 \draw [dashed] (0,2) -- (1,2);
 \draw [dashed] (3,2) -- (4,2);
 \draw[red,->] (1,2) -- (1,3) ; 
 \draw[red,->] (3,3) -- (3,2) ;
 \draw[olive,dotted] (2,1.5) -- (1,2) ;
 \draw[olive,dotted] (2,1.5) -- (3,2) ;
    \end{tikzpicture}
\hspace{0.1\columnwidth}
%
%
   \begin{tikzpicture}[scale=0.8,very thick]
\node (lab) at (3.8,2.5) {\textbf{\textit{(f)}}} ;
\node [red] (A) at (1.3,2.6) {$\vec{p}$} ;
\node [red] (B) at (2.7,2.6) {$\vec{p}$} ; 
\node [red] (AB) at (2.0,3.3) {$\vec{p}+\vec{q}$} ;
\node [blue] (BB) at (0.3,3.3) {$\vec{q}$} ;
\node [blue] (CC) at (3.7,3.3) {$\vec{q}$} ;
\node (N1) at (0.1,2.2) {N$^{\phantom{\prime}}$} ;
\node (N2) at (0.1,1.7) {N$^{\prime}$} ;
 \draw[cyan,fill=cyan] (0,0) -- (4,0) -- (4,1) -- (0,1) --cycle  ;
 \node (DD) at (1.9,0.4) {\large{A-2}} ;
 \draw [] (0,1.5) -- (4,1.5);
 \draw [] (0,2) -- (1,2);
 \draw [] (3,2) -- (4,2);
 \draw[red,->] (1,2) -- (1,3) ; 
 \draw[red,->] (3,3) -- (3,2) ;
 \draw[red,->] (1,3) -- (3,3) ;
 \draw[blue,->,dotted] (0.5,3.5) -- (1,3) ;
 \draw[blue,->,dotted] (3,3) -- (3.5,3.5) ;
    \end{tikzpicture}
    \caption{Schematic representation of the single-nucleon momentum distribution $n^{[1]}(p)$ of Eq.~(\ref{eq:NMD}) ((a)) and how it is computed in the LCA ((b)-(e)). The black dashed lines denote IPM nucleons. The dotted olive lines denote one of the correlation operators in Eq.~(\ref{eq:SRCoperator}). Diagram (b) is the IPM contribution and dominates for $p<p_F$. Diagrams (c), (d) and (e) are SRC corrections with $NN^{\prime}$ pairs and are the major contributors to $n^{[1]}(p)$  for  $p>p_F$. Diagram (f) highlights the connection between SRC sensitive $A(e,e'N(N^{\prime}))$ measurements with momentum transfer $\vec{q}$ and the fat tail of $n^{[1]}(p)$: the experimental cuts on events with $\mid \vec{p} \mid > p_F$ and low excitation energies for $A-2$ selects the reaction mechanisms of diagrams (c)-(e).}
    \label{fig:schememomendis}
\end{figure}
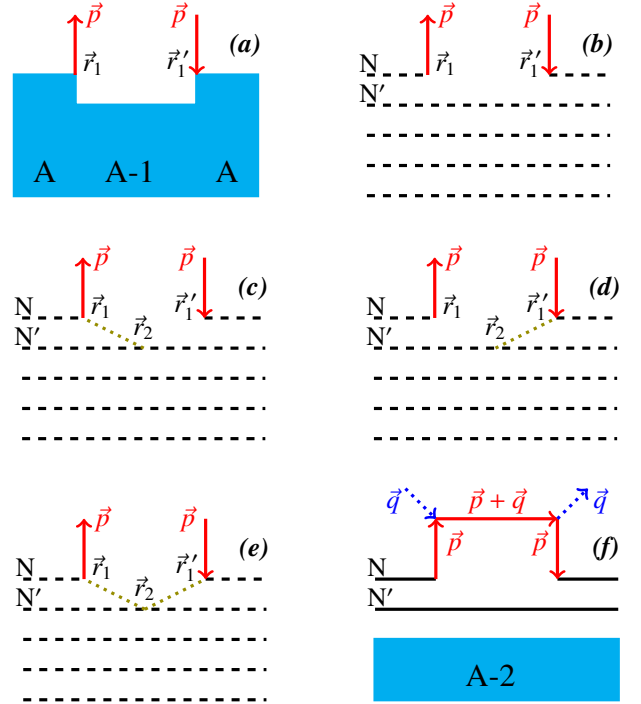

\section{Framework and results}
\label{sec:results}
The single-nucleon momentum distribution $n^{[1]}(p)$ for nucleus $A$ is defined as
\begin{eqnarray} \label{eq:NMD}
  n^{[1]}(p) & = & \frac{A}{(2\pi)^3}
  \int d^2 \Omega_p  
  \int d^3 \vec{r}_1\;
   d^3 \vec{r}_1^{\,\prime}\;
   d^{3(A-1)} \{\vec{r}_{2-A} \} 
  \nonumber \\
  & & 
  \times e^{-i \vec{p}\cdot(\vec{r}^{\,\prime}_1 - \vec{r}_1)} \; 
   \Psi^*(\vec{r}_1,\vec{r}_{2-A})
  \Psi(\vec{r}_1^{\,\prime},\vec{r}_{2-A}) \; ,
\end{eqnarray}
where $\vec{r}_{2-A} \equiv \vec{r}_2 \vec{r}_3 \ldots \vec{r}_A$, $\Omega_p$ is the angular part of $\vec{p}$ and $\Psi$ is the ground-state wave function of the nucleus under study. In Ref.~\cite{Vanhalst:2014cqa}, a methodology to compute the SRC contribution to $n^{[1]}(p)$ was developed. In a nutshell, the  complexity of the calculation is shifted from the wave functions to the operators
\begin{equation}
\left| \Psi \right> = 
\frac{1}
{\sqrt{\left<\Phi \right| \widehat{\mathcal{G}}^{\dagger} \widehat{\mathcal{G}} \left| \Phi \right>   }} \widehat{\mathcal{G}} \left|\Phi \right> \; ,
\end{equation}
where $\left|\Phi \right>$ is a Slater determinant reminiscent of the IPM. Several calculations have indicated that the major sources of SRC correlations  are the central (Jastrow), tensor and
  spin-isospin terms. Accordingly, in our approach the nuclear SRC  operator $\widehat{\mathcal{G}}$ adopts the form 
\begin{eqnarray} \label{eq:SRCoperator}
\widehat{\mathcal{G}} & = &   \widehat{{\cal S}}  
\biggl( \prod _{i<j=1} ^{A} \biggl[ 1  
-  {g_c(r_{ij})} 
+  {f_{t\tau}(r_{ij}) } \widehat{S}_{ij} \vec{\tau}_i \cdot 
  \vec{\tau}_j \; 
\nonumber \\
& & + 
 { f_{\sigma \tau}(r_{ij}) } \vec{\sigma}_i \cdot \vec{\sigma}_j 
\vec{\tau}_i \cdot \vec{\tau}_j 
\biggr] \biggr)  \; ,
\end{eqnarray}
where $\widehat {{\cal S}}$ and $\widehat{S}_{ij}$ are the symmetrization and tensor operator. 
 In LCA, a truncation procedure is adopted and the SRC-induced corrections to  $n^{[1]}(p) $ are two-body operators whereby all terms that are linear and quadratic in the correlation functions are retained (see Fig.~\ref{fig:schememomendis}). Thereby, the normalization condition
 can be preserved. We assume that short-distance operators $\widehat{\mathcal{G}}$ of the type (\ref{eq:SRCoperator}) between IPM nucleons can account for high-momentum inter-nucleon correlations~\cite{Bogner:2012zm}. We use a set of correlation functions that we have systematically tested in comparisons of reaction-model calculations and SRC-driven data 
\cite{Blomqvist:1998gq, Onderwater:1998zz,Starink:2000qhh, Colle:2015ena}. The $f_{t 
\tau}(r)$ and $f_{\sigma \tau}(r)$ correlation functions are from a variational 
calculation~\cite{Pieper:1992gr}.  
An analysis of $^{12}$C$(e,e^{\prime}pp)$~\cite{Blomqvist:1998gq} and  $^{16}$O$(e,e^{\prime}pp)$~\cite{Starink:2000qhh} experimental results  systematically excluded ``soft'' central correlation functions $g_c$ and preferred  a ``hard'' $g_c$ inferred from a G-matrix calculation with the Reid soft-core interaction in nuclear matter~\cite{Dickhoff:2004xx}. Different interactions generate different correlations --particularly for the central correlations-- and are sources of theoretical uncertainties. In order to quantify those and to provide some sense of the robustness of our calculations, we present LCA results with the ``hard'' $g_c$ from~\cite{Dickhoff:2004xx} and the ``soft'' $g_c$ from \cite{Pieper:1992gr} that is consistent with the adopted $f_{t 
\tau}$ and $f_{\sigma \tau}$.  We use harmonic oscillator (HO) single-particle states as they offer the 
possibility to separate the pair's relative and c.m.~motion with the aid of 
Moshinsky brackets. As the major purpose of this study is to determine the systematic properties, we use the HO parameters from the global parameterization $\hbar \omega=45{A}^{-\frac {1} {3}} - 25{A}^{-\frac {2} {3}} $. More advanced calculations could find the optimum HO parameter for each specific nucleus but this complication is beyond the scope of the current paper. It has been numerically shown~\cite{ Vanhalst:2014cqa, Ryckebusch:1997gn, Colle:2015lyl,   Ryckebusch:1996wc, Colle:2013nna} and experimentally confirmed~\cite{Starink:2000qhh, Colle:2015ena} that the major source of SRC strength 
stems from correlation operators acting on IPM pairs in a nodeless relative $S$ state. This can be intuitively understood by noting that the 
probability of finding close-proximity IPM pairs is dominated by pairs in a nodeless 
relative $S$ state. Those wave functions are not very sensitive to the details of the mean-field potential.

\begin{figure}[ht]
\centering
\includegraphics[viewport=20 19 527 399, clip, width=0.95\columnwidth]{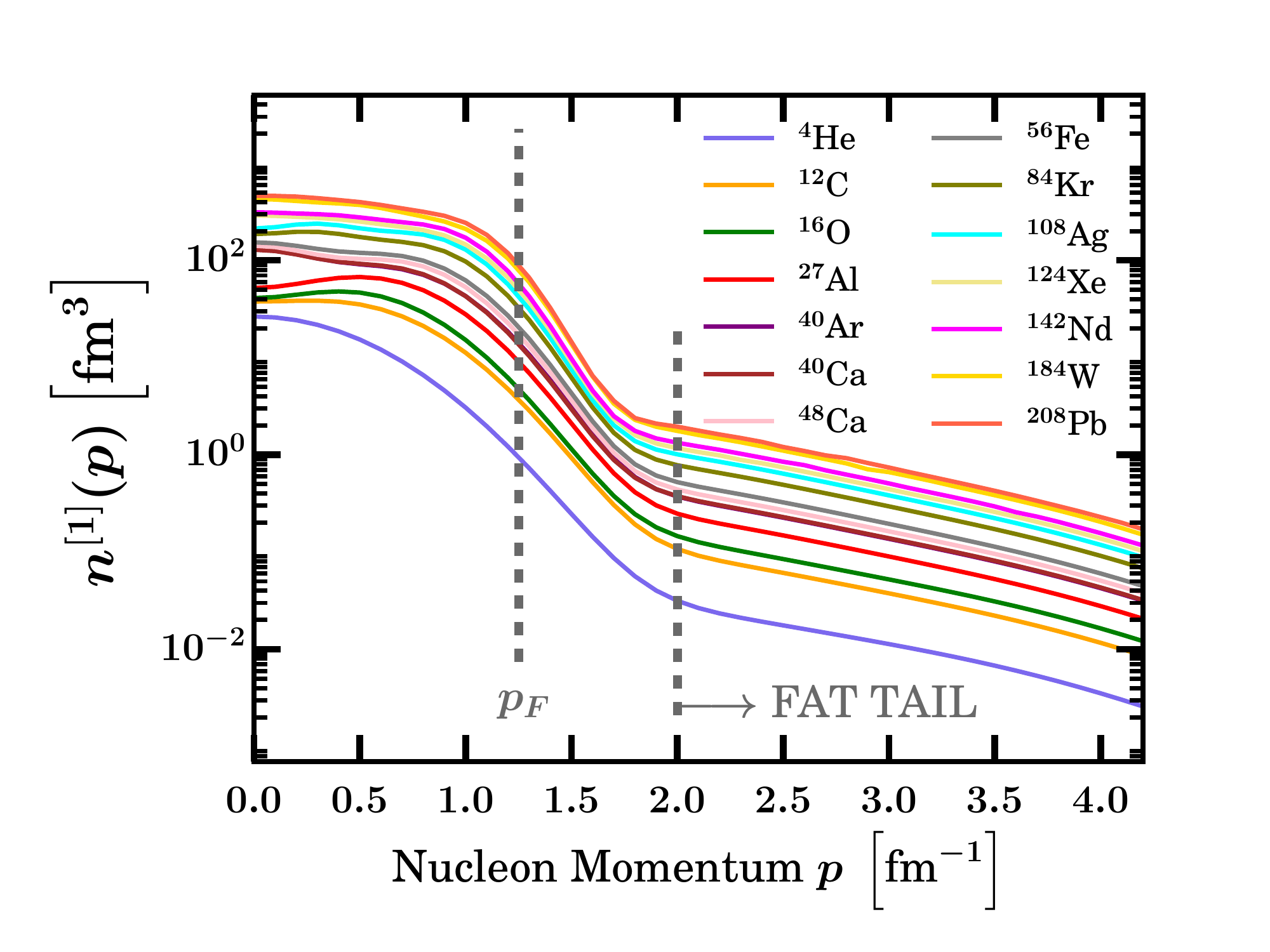}
\caption{The  momentum distribution for 14 nuclei across the nuclear mass table. The $n^{[1]}(p)$ are computed in LCA with a ``hard'' central correlation function $g_c$ adopting the normalization convention $\int dp \; p^{2} n^{[1]}(p) =A$.} 
\label{figmomentumdis}
\end{figure}

\begin{figure}[ht!]
\centering
\includegraphics[width=0.50\columnwidth, viewport=46 78 534 411,clip]{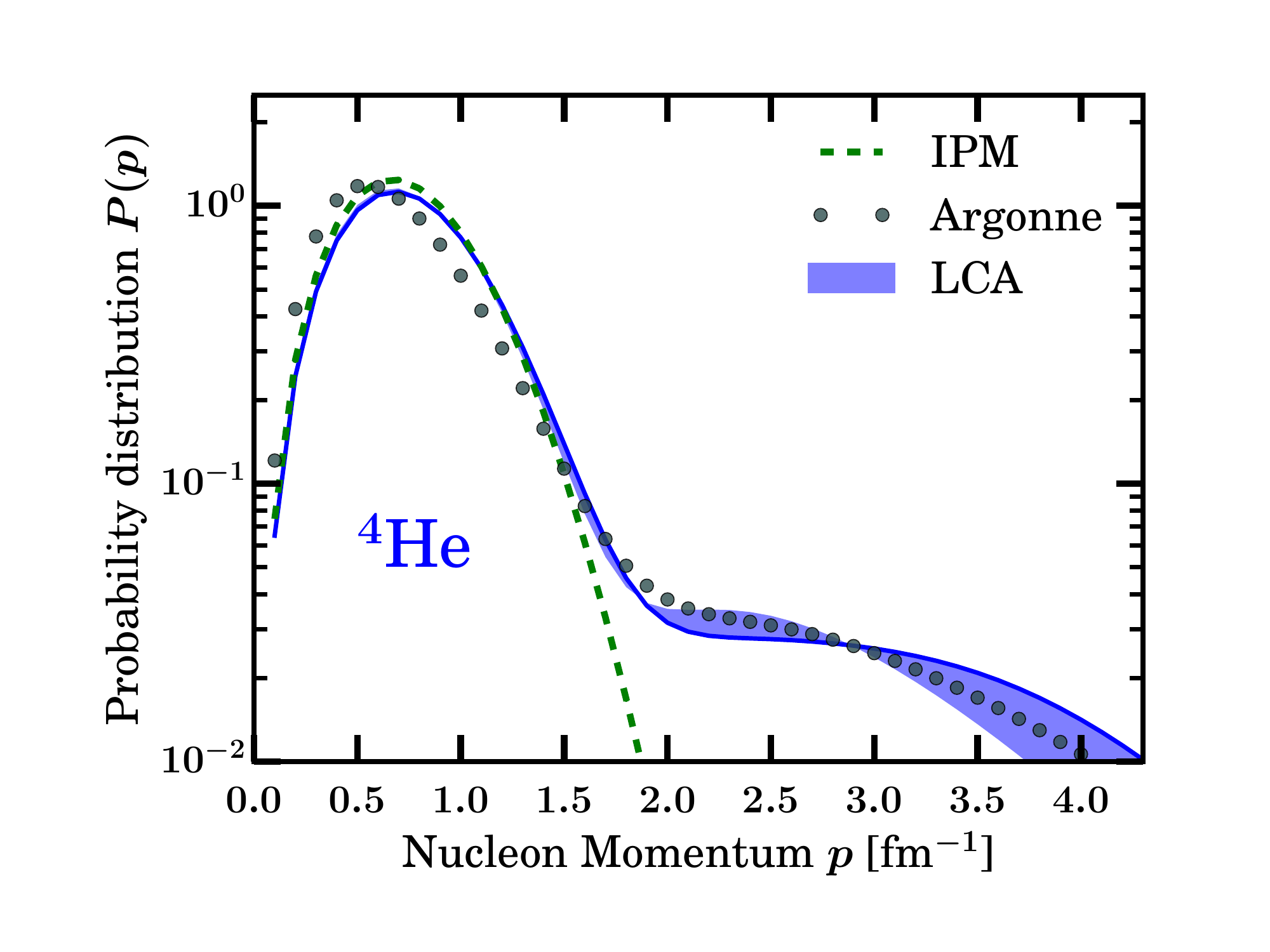}
 \includegraphics[width=0.4754125\columnwidth, viewport=70 78 534 411,clip]{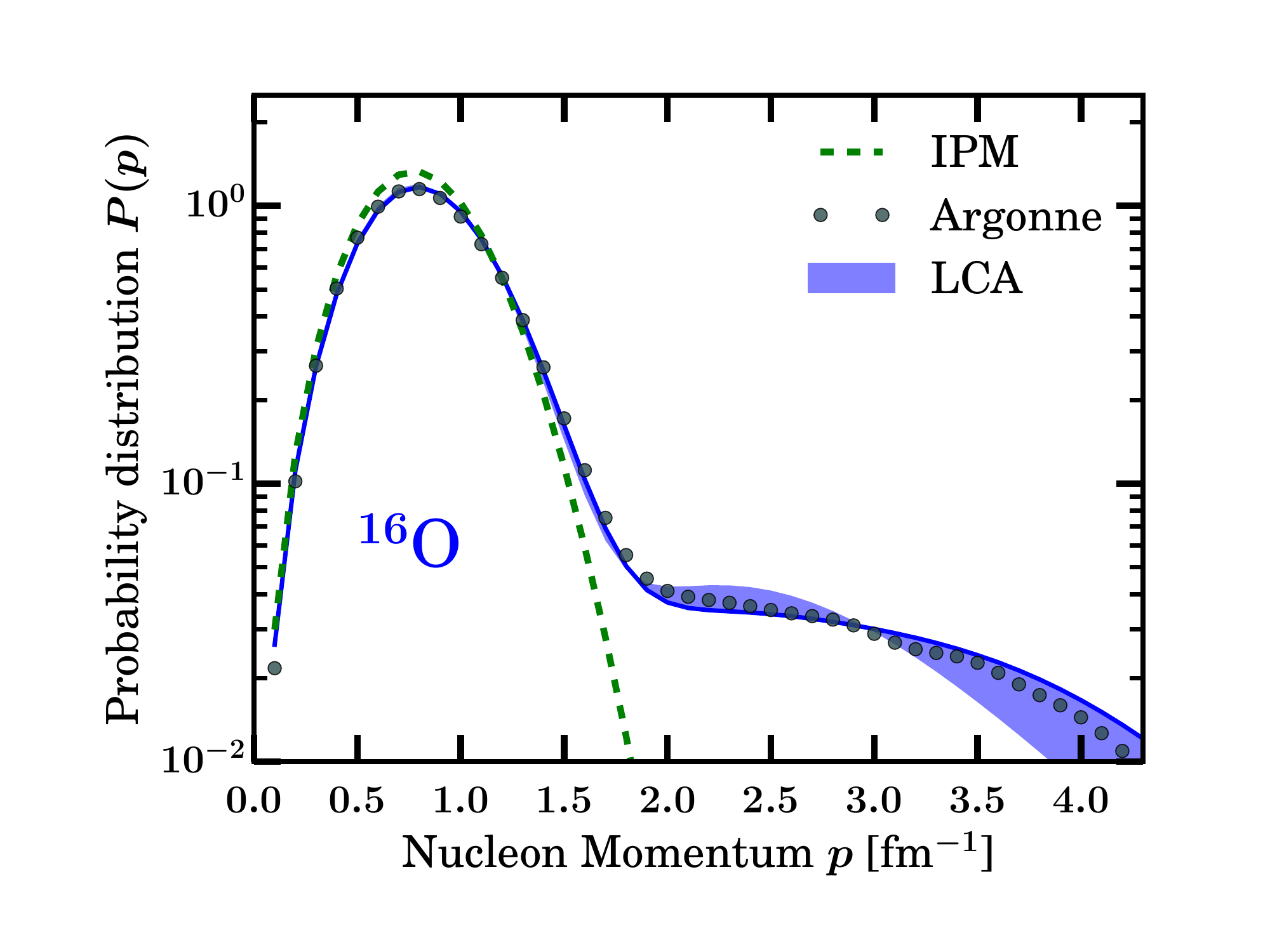}
 \includegraphics[width=0.50\columnwidth, viewport=46 78 534 411,clip]{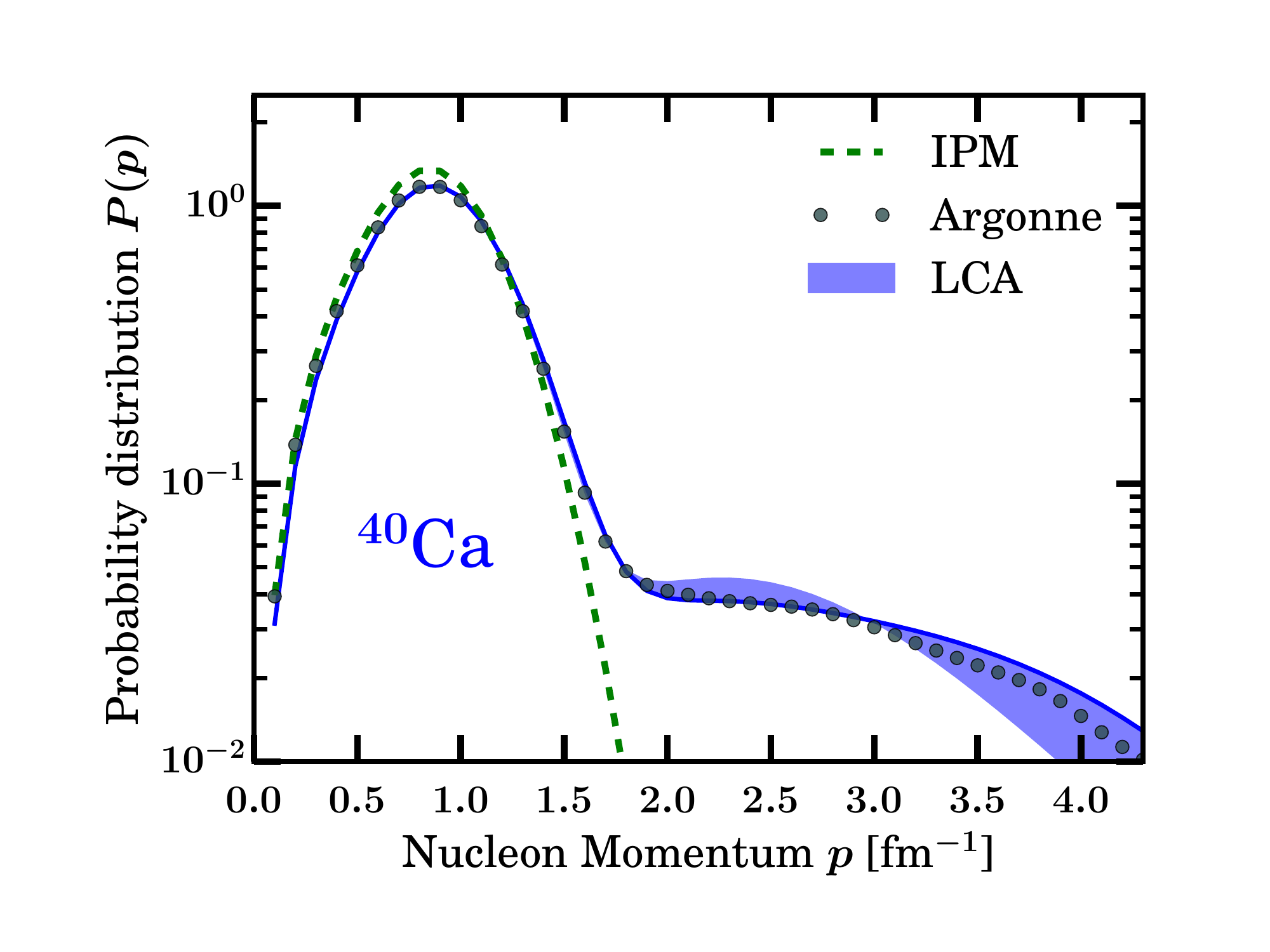}
 \includegraphics[width=0.4754125\columnwidth, viewport=70 78 534 411,clip]{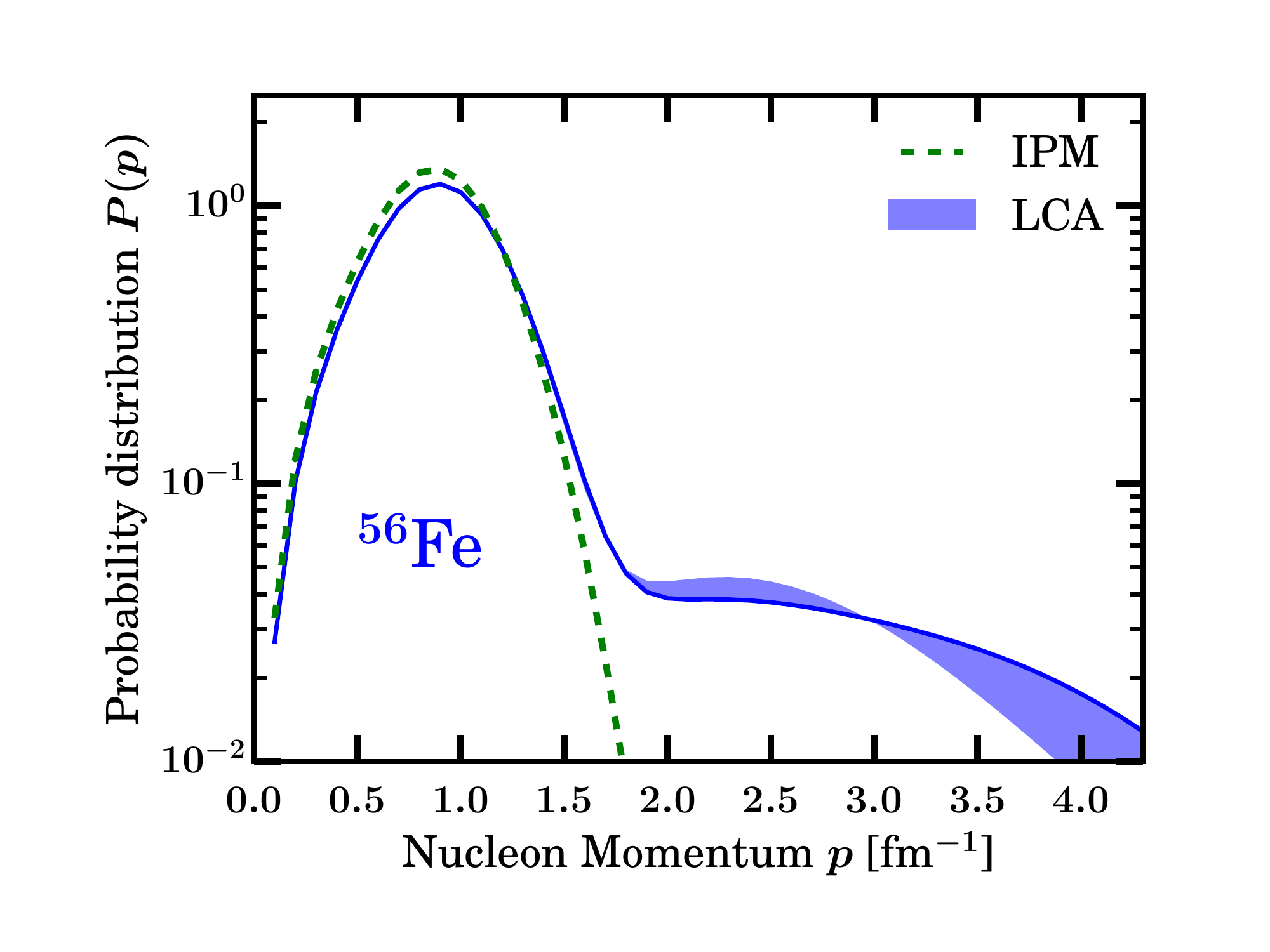}
\includegraphics[width=0.50\columnwidth, viewport=46 28 534 411,clip]{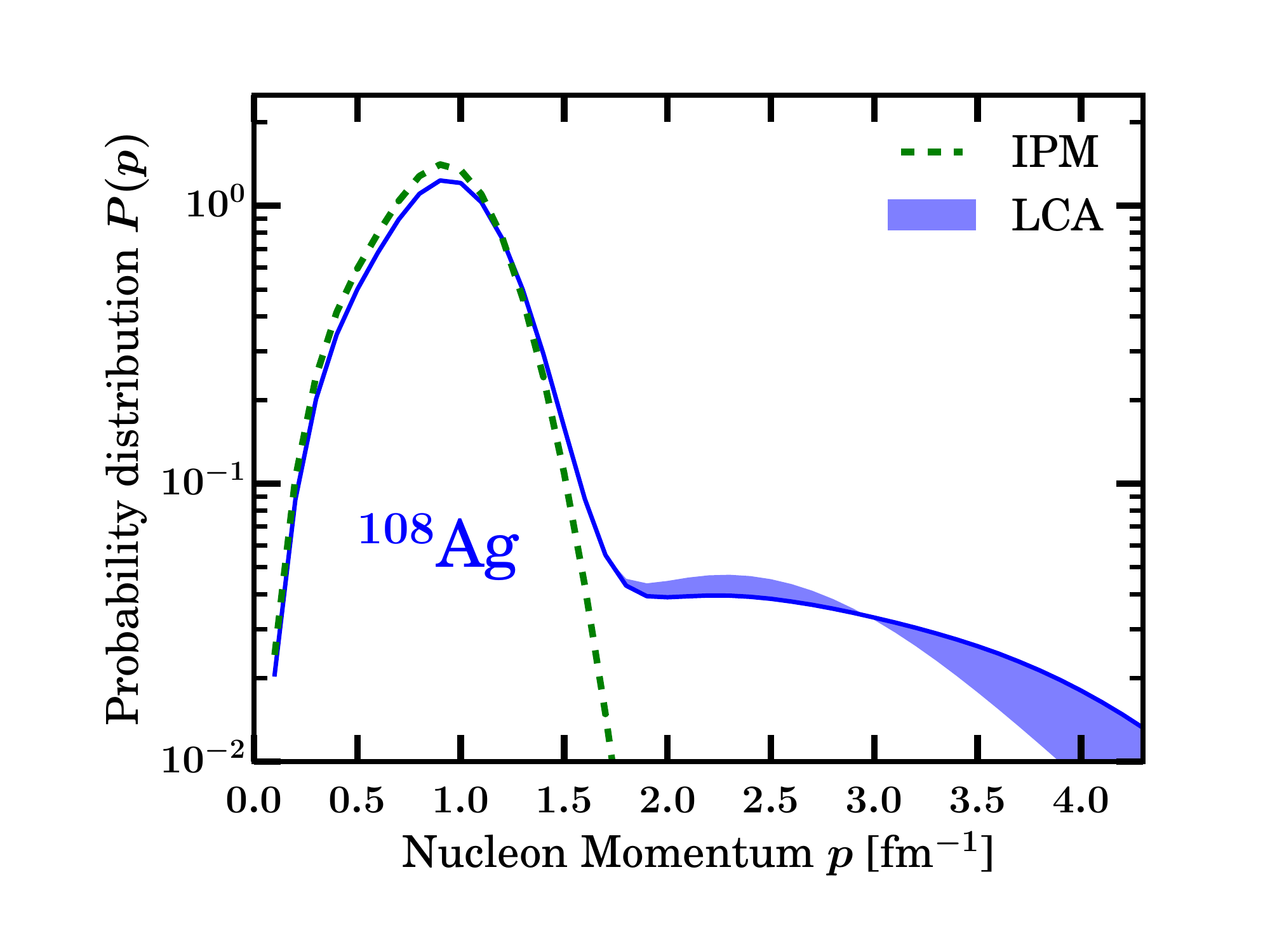}
 \includegraphics[width=0.4754125\columnwidth, viewport=70 28 534 411,clip]{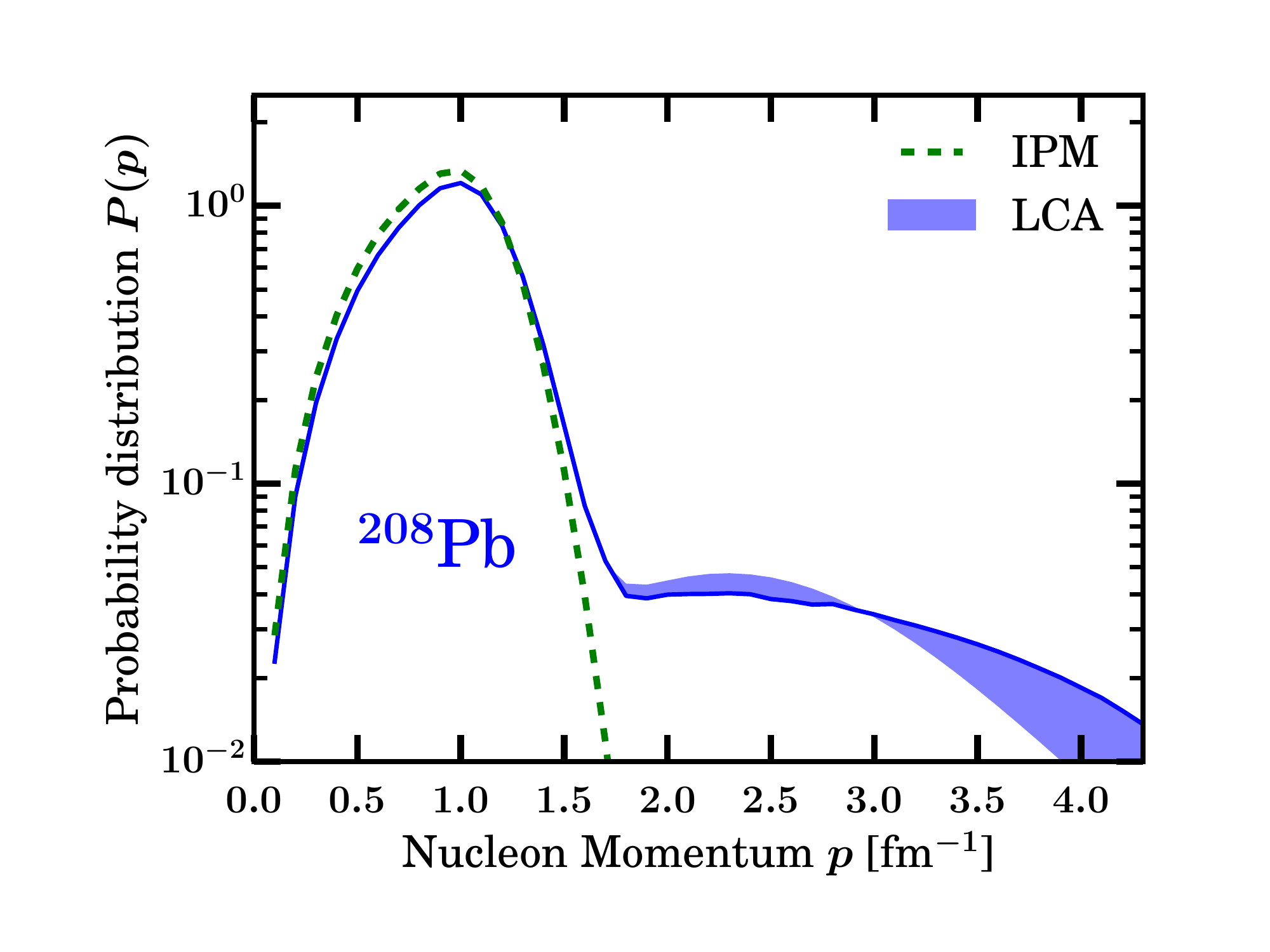}
\caption{The probability distribution $P(p)$ to find a nucleon with momentum $p$ in six  nuclei. The IPM and LCA results are shown, as well as the QMC results (when available) from the Argonne group with the AV18 nucleon-nucleon interaction.The LCA bands indicate the difference between the calculations with a ``soft'' and a ``hard'' central correlation function $g_c$. The solid line is the LCA result with a ``hard'' $g_c$.} 
\label{figprobandcdf}
\end{figure}

\begin{figure}[ht]
\centering
\includegraphics[width=0.50\columnwidth, viewport=46 75 534 411,clip]{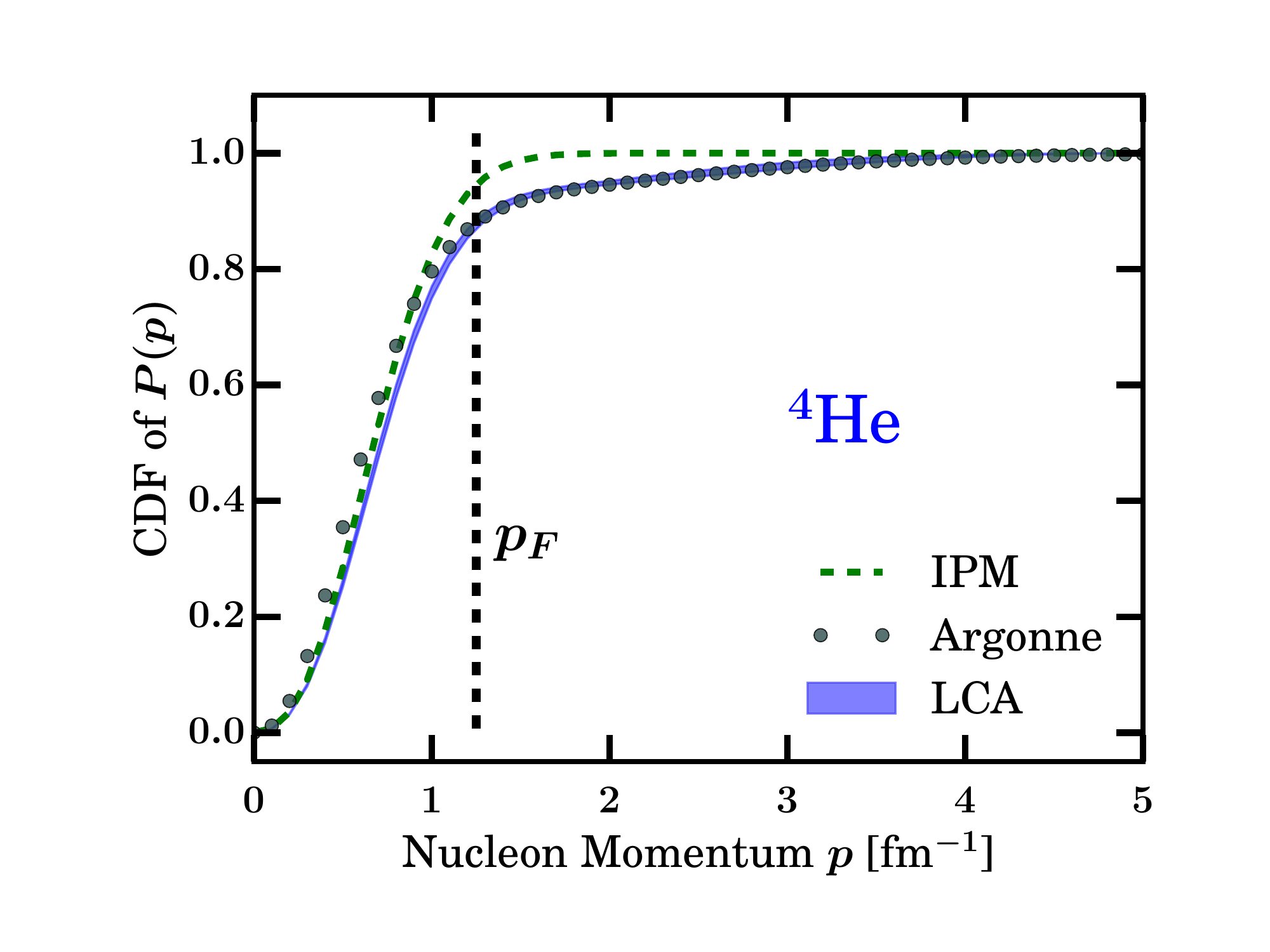}
 \includegraphics[width=0.463115\columnwidth, viewport=82 75 534 411,clip]{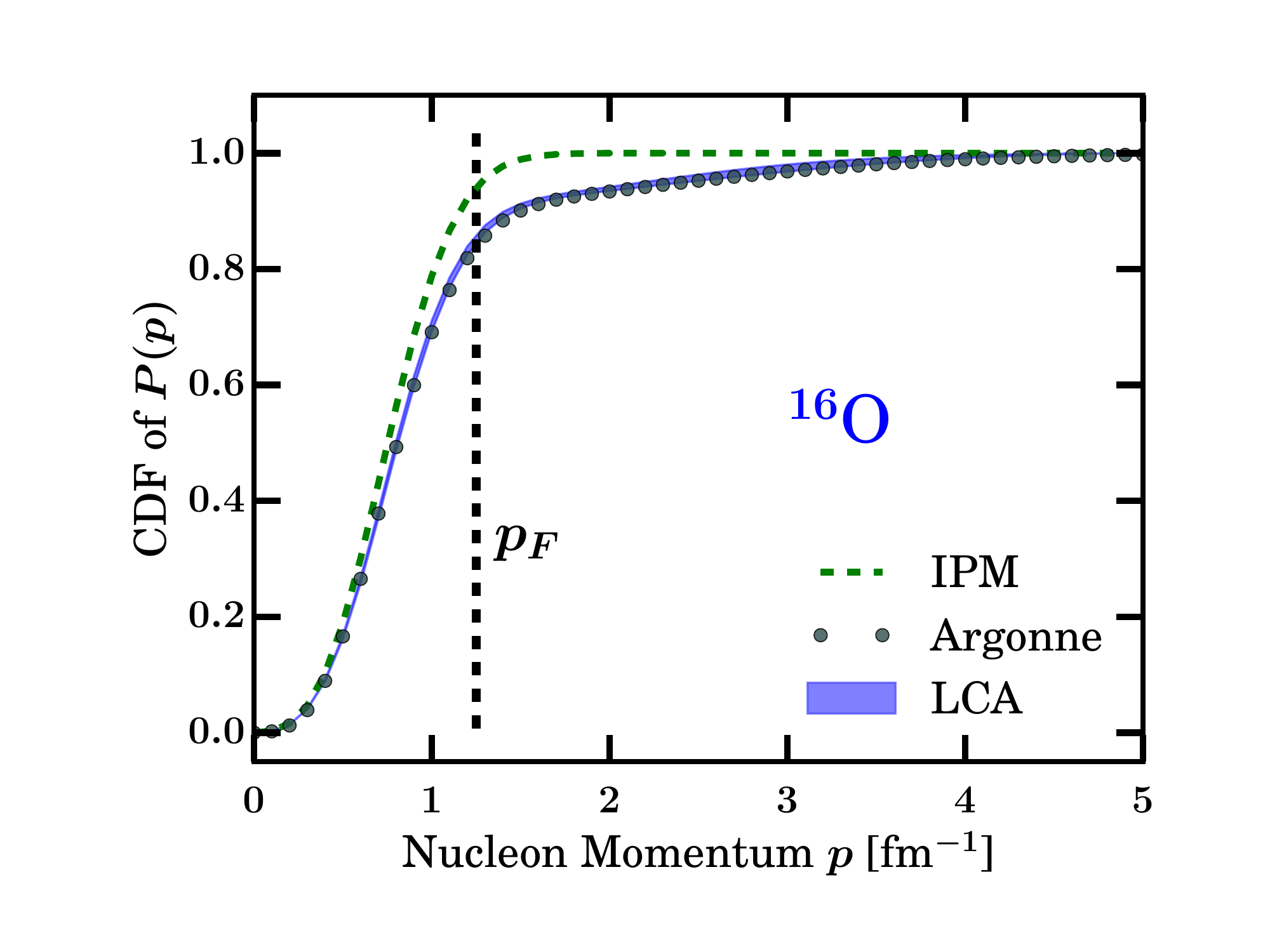}
\includegraphics[width=0.50\columnwidth, viewport=46 28 534 411,clip]{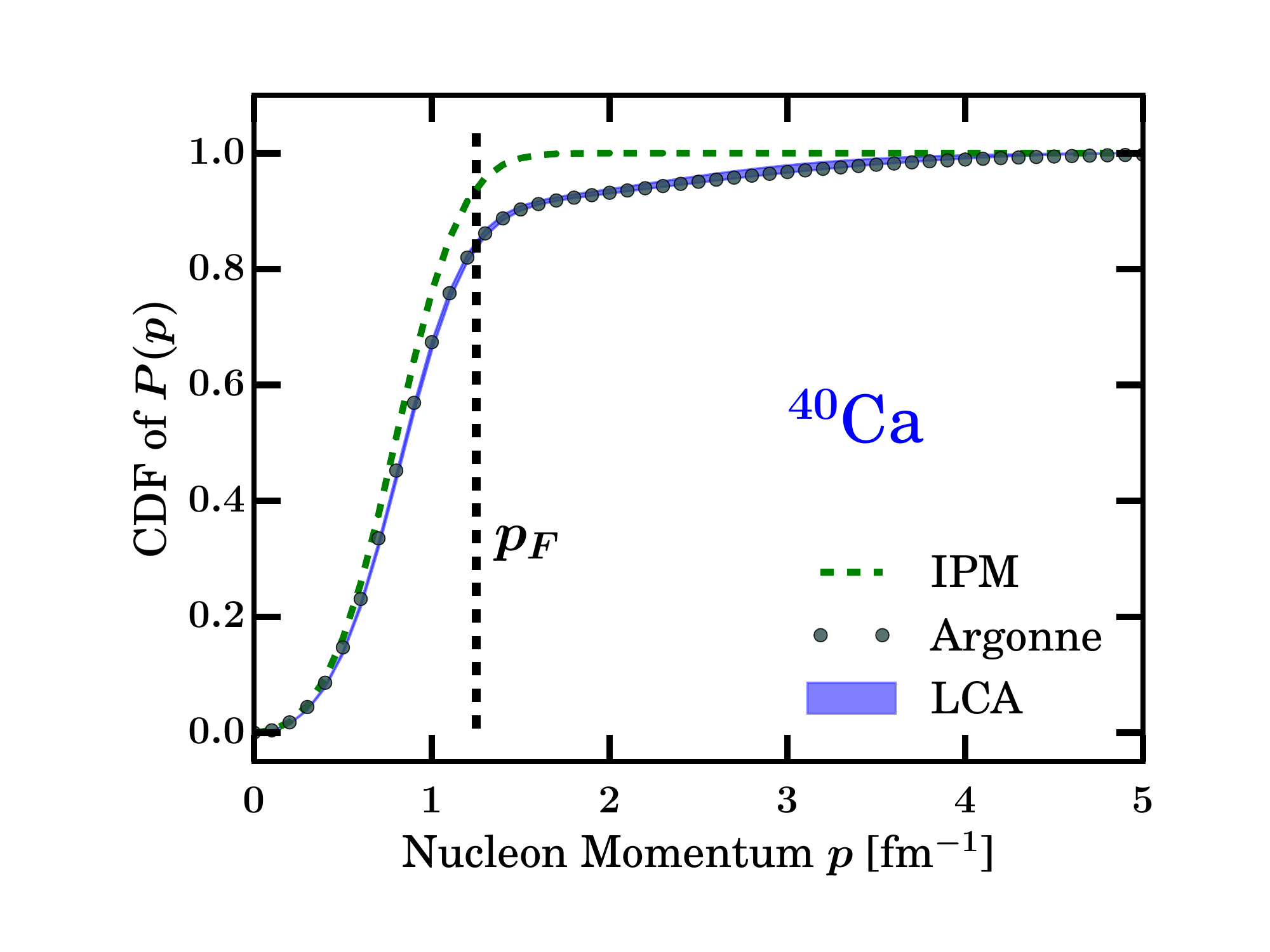}
 \includegraphics[width=0.463115\columnwidth, viewport=82 28 534 411,clip]{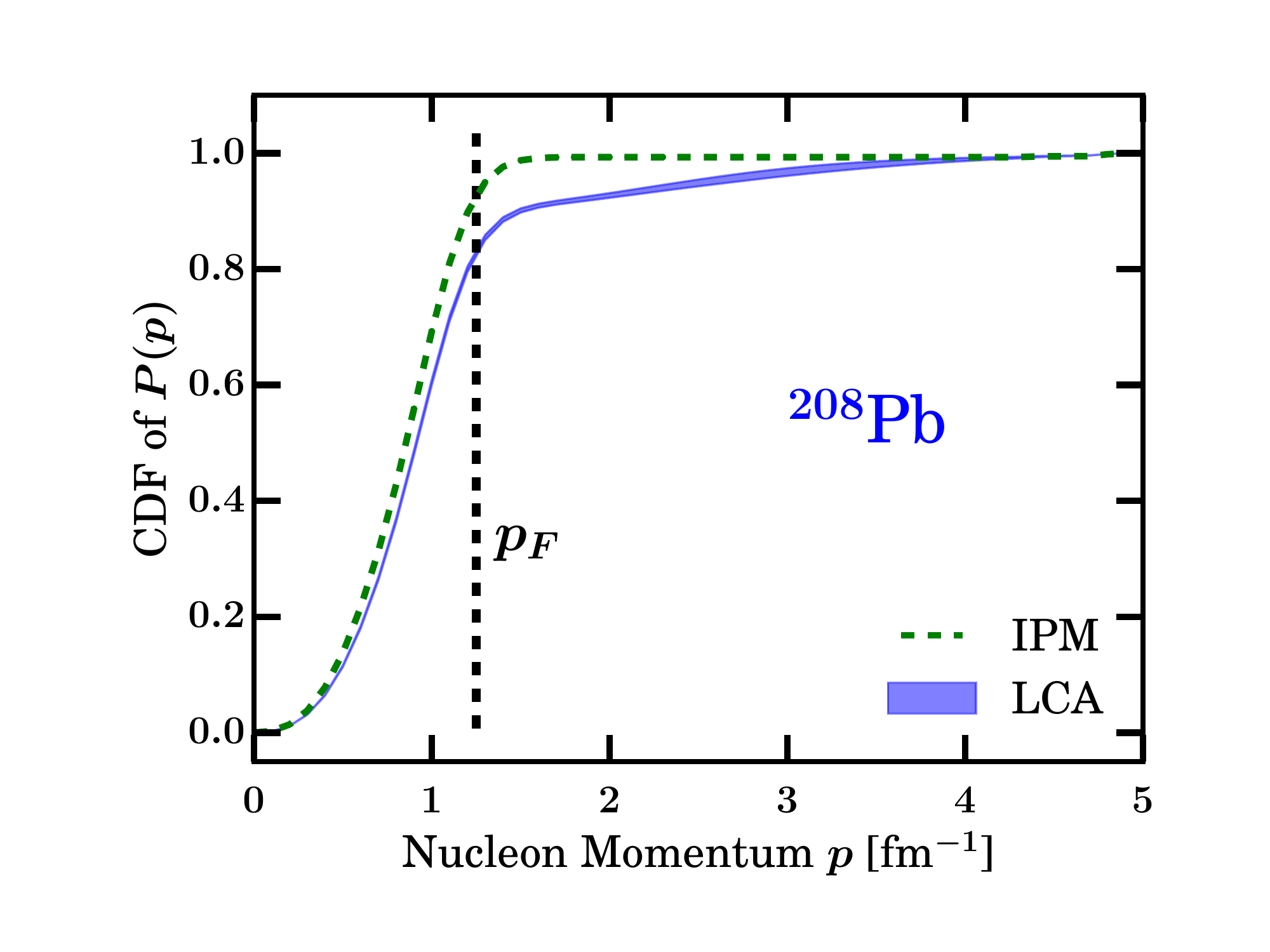}
\caption{As in Fig.~\ref{figprobandcdf} but for the cumulative distribution function (CDF) for finding a nucleon with momentum $p$. } 
\label{figcdf}
\end{figure}

We involve 14 nuclei in the presented study: four symmetric ones $\left(  ^{4} \text{He}, ^{12} \text{C}, ^{16} \text{O}, ^{40} \text{Ca}   \right)$ and ten asymmetric ones, covering a neutron excess range $\frac{N}{Z}$ from 1.0 up to 1.54 (for $^{208}$Pb).  Figure~\ref{figmomentumdis} displays the momentum dependence of the  $n^{[1]}(p)$ as computed in the LCA. The results illustrate that a clear scale separation between a ``low'' and ``high'' nucleon momentum  regime can be made. For the heavier nuclei there is some distinct intermediate momentum regime $1.25 \lesssim p \lesssim 2$~fm$^{-1}$. The IPM part is defined by momenta smaller than the Fermi momentum $ p_F=1.25~$fm$^{-1}$ and is characterized by a clear nuclear mass dependence. From $^{12}$C onward, the $n^{[1]}(p<p_F)$ increasingly resembles the one for a Fermi gas with growing mass number $A$. At nucleon momenta $p \gtrsim 2$~fm$^{-1}$ the $n^{[1]}(p)$ are fat-tailed and this characterizes the SRC part of the momentum distribution. Across all nuclei the tails of the momentum distribution are just scaled versions of one another. One can infer that the same generating mechanism is at play across all nuclei. In Ref.~\cite{Vanhalst:2014cqa}, we identified this mechanism as correlation operators acting on IPM pairs in a nodeless relative $S$-wave that account for about 90\% of the strength in the correlated part of the $n^{[1]}(p)$.   The observation that the momentum dependence of the fat-tailed part of $n^{[1]}(p)$ is universal (modulo a scaling factor) is often referred to  as the universality of SRC~\cite{Rios:2013zqa, Alvioli:2016wwp, Feldmeier:2011qy,   Weiss:2016obx, Artiles:2016akj, Mosel:2016uge, Frankfurt:1981mk}. From Fig.~\ref{figmomentumdis} we infer that the fat tail of the single-nucleon momentum distributions decreases exponentially with $p$, a feature that can be understood on general grounds~\cite{PhysRevC.14.1264}. 

Of great relevance for reactions involving nuclear targets is the probability distribution $P(p) = p^2 n^{[1]}(p)/A$ to find a nucleon with momentum $p$.   Figure~\ref{figprobandcdf} shows the computed $P(p)$ for six selected nuclei. Below $p_F$, the   $P(p)$ is Maxwell-Boltzmann like. Obviously, the SRC induce fat tails to the $P(p)$ and the probability of finding a very fast nucleon is many orders of magnitude larger after including SRC. The SRC also reduce the probability of finding a ``slow'' nucleon. We compare the LCA results for $P(p)$ with those from ab-initio Quantum Monte-Carlo (QMC) calculations by the Argonne group~\cite{Wiringa:2013ala, Lonardoni:2017egu}.  Not surprisingly, the strongest deviation between the LCA and Argonne results are observed for the light nucleus $^{4}$He. For $^{16}$O and $^{40}$Ca (the heaviest nucleus for which QMC results are available), the QMC results are compatible with the LCA ones given the theoretical uncertainties. Large theoretical uncertainties related to the choice of $g_c$ occur for $p \gtrsim 3$~fm$^{-1}$. Thereby, the LCA results with the ``hard'' $g_c$ reproduce the QMC results better than those with the ``soft'' $g_c$.   We stress that LCA does not account for long-range correlations that are an additional source of strength in the low-momentum part of the fat tail ($p \approx 2$~fm$^{-1}$). We conclude that all in all a fair representation of the momentum dependence of the probability distributions can be obtained in a model that corrects the IPM for SRC between pairs of nucleons. This is further illustrated with the cumulative distribution functions (CDF) of $P(p)$ in Fig.~\ref{figcdf} for which we find relatively minor theoretical uncertainties related to the choice for $g_c$.  For lead, about 80\% of the nucleons have a momentum lower than the Fermi momentum. For oxygen and calcium, the LCA and QMC CDFs are comparable. Note that the $p \gtrsim 3$~fm$^{-1}$ range, for which relatively large theoretical uncertainties emerge in the $P(p)$ of Fig.~\ref{figprobandcdf}, represents but a few percent of the total amount of nucleons.

\begin{figure}[htb]
    \centering
    \includegraphics[width=0.50 \columnwidth, viewport=54 78 534 411,clip]{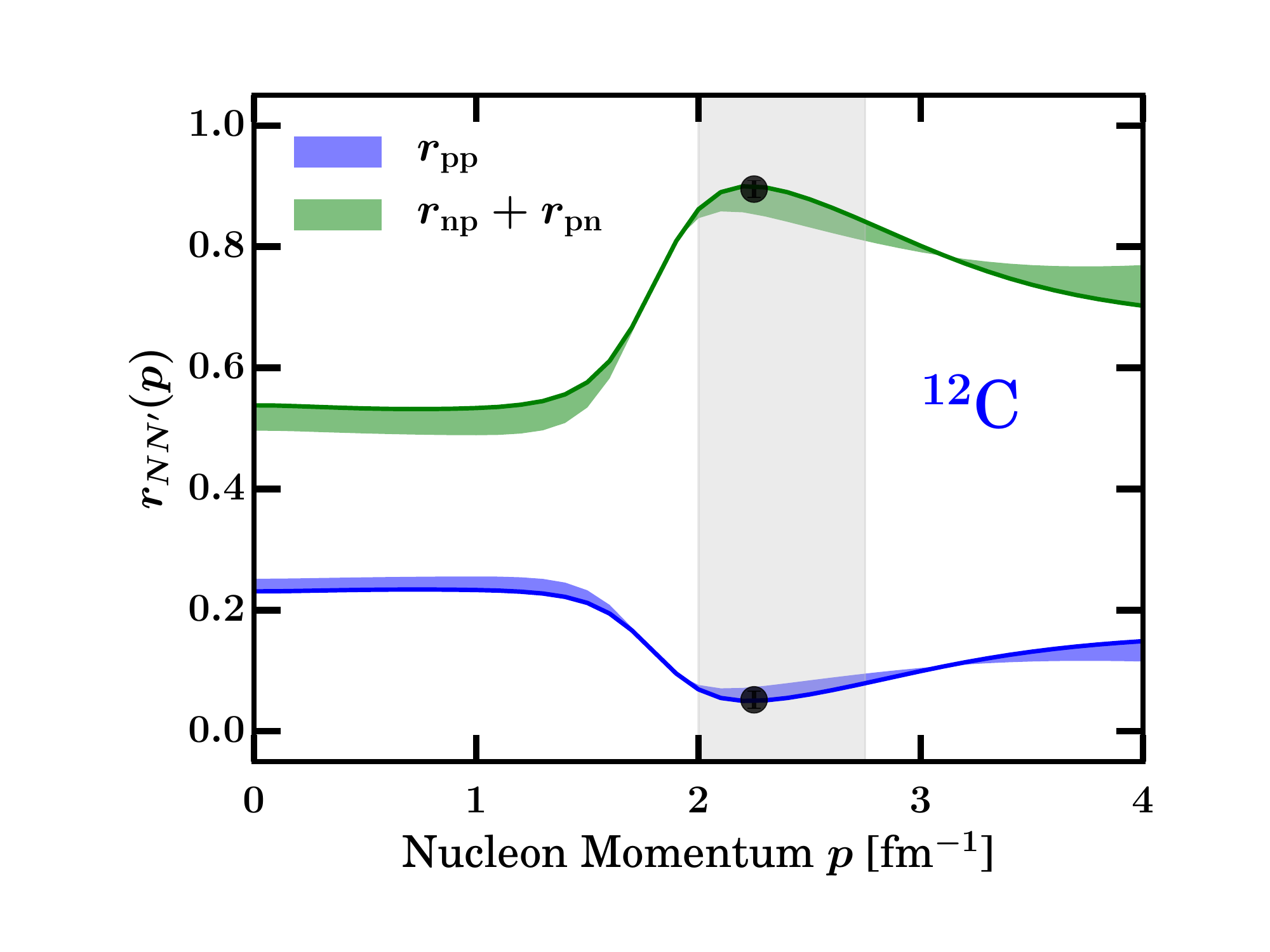}
    \includegraphics[width=0.47396 \columnwidth, viewport=79 78 534 411,clip]{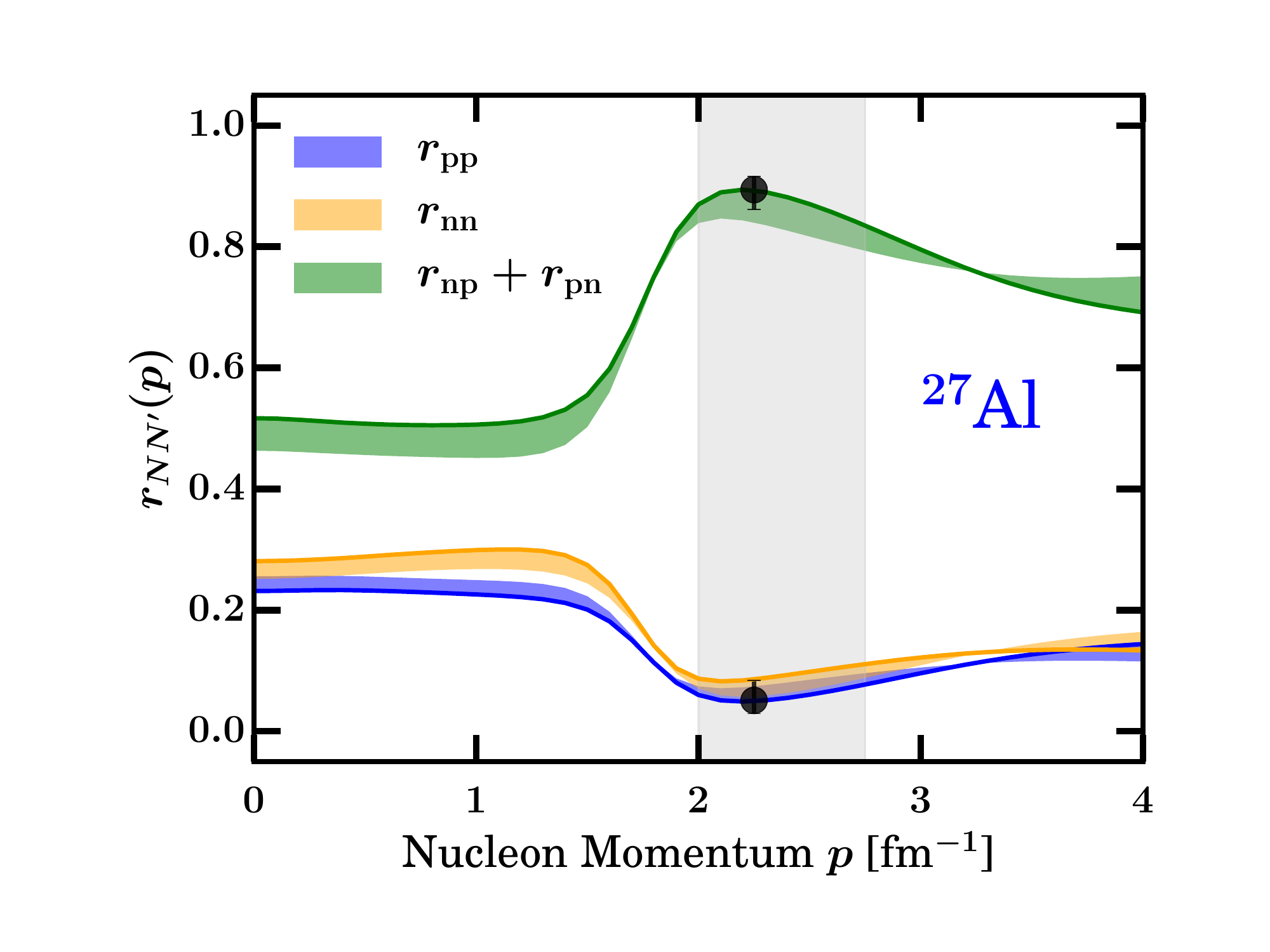}
    \includegraphics[width=0.50 \columnwidth, viewport=54 28 534 411,clip]{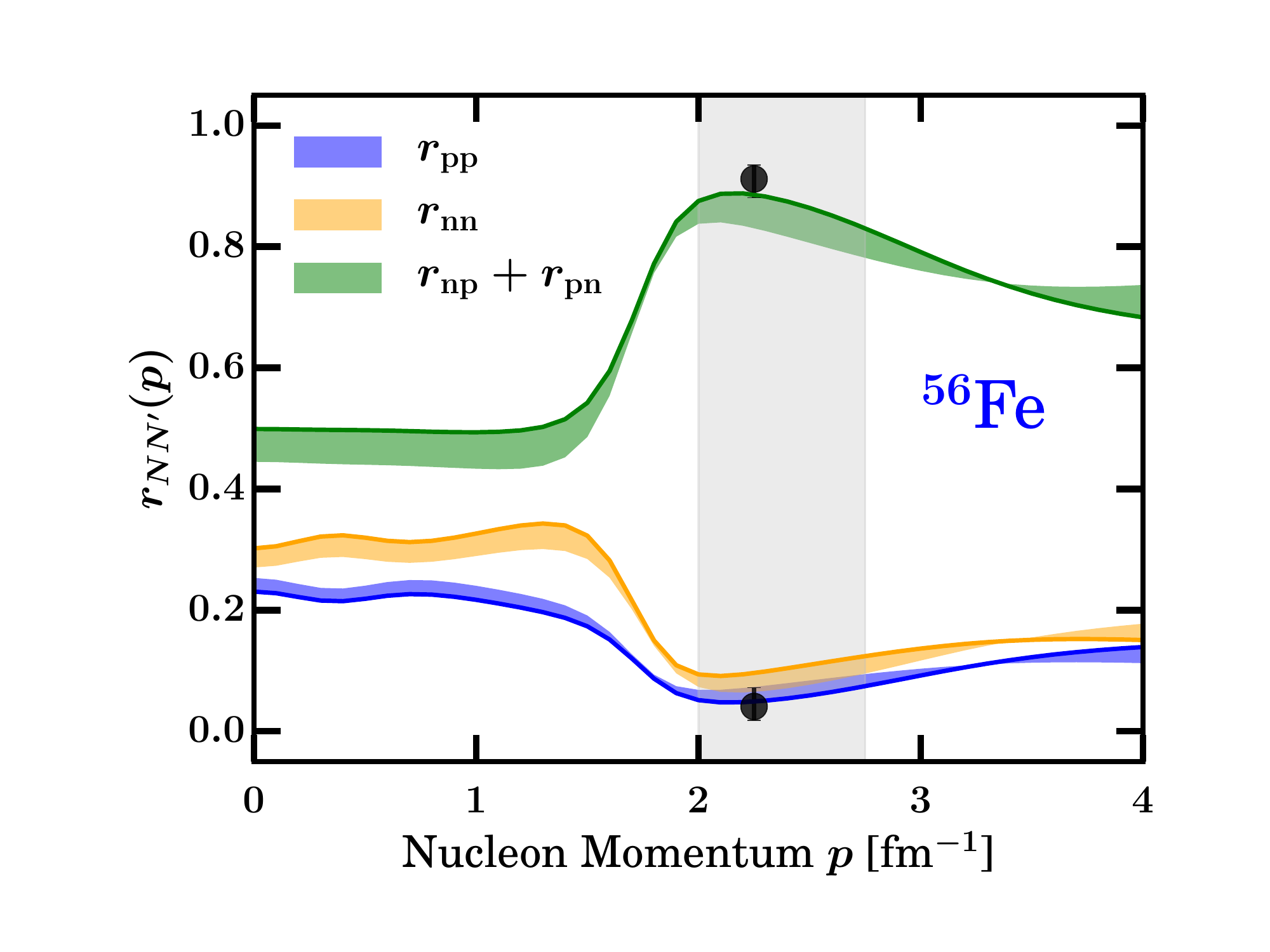}
    \includegraphics[width=0.47396 \columnwidth, viewport=79 28 534 411,clip]{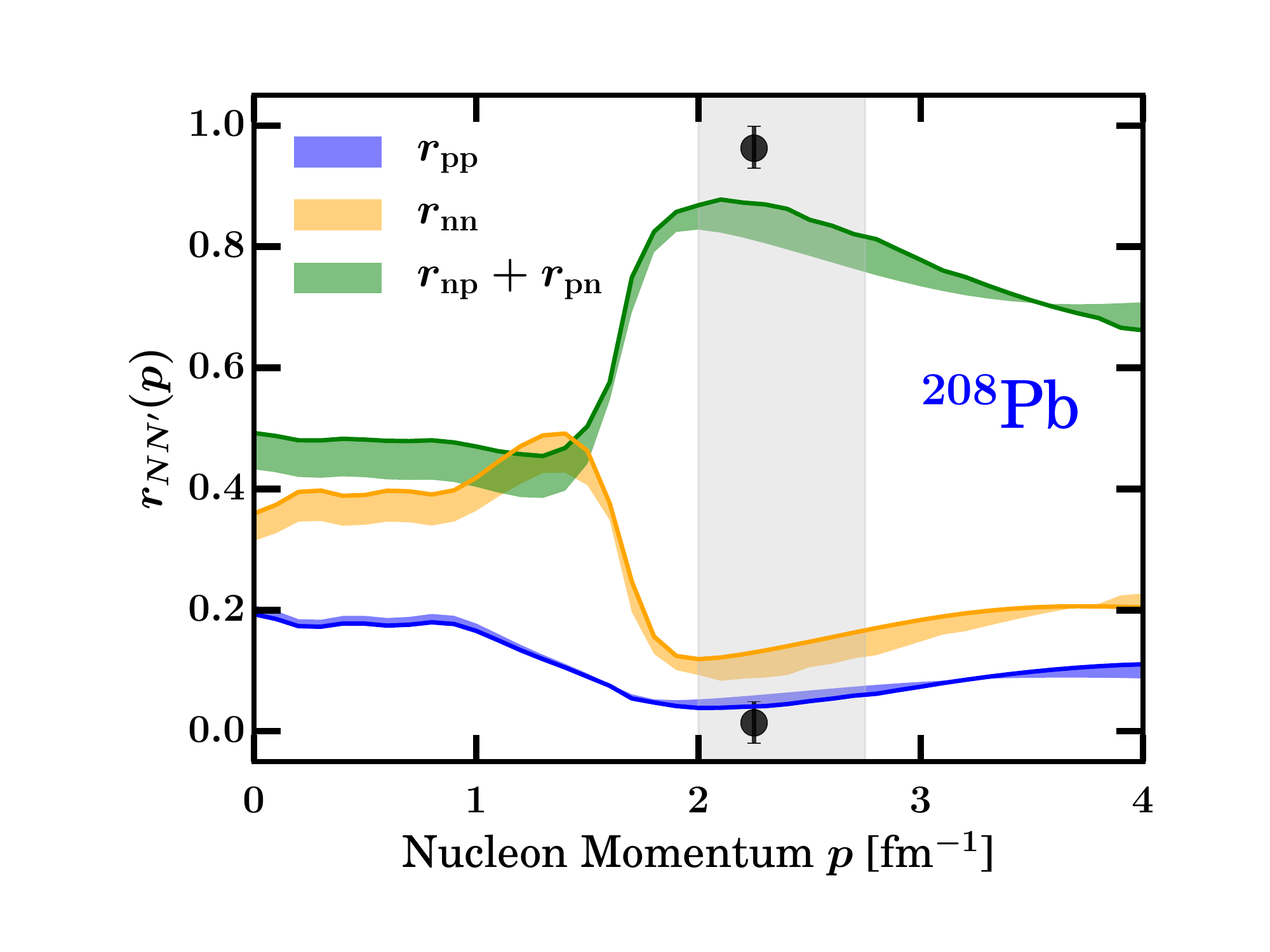}
    \caption{Pair composition of the single-nucleon momentum distribution as computed in LCA. The bands quantify the theoretical uncertainty related to the choice of the $g_c$. The solid lines are obtained with the hard ``$g_c$''. For the $N=Z$ nucleus $^{12}$C the $r_{\text{pp}}$ and $r_{\text{nn}}$ coincide. The data are from Ref.~\cite{Hen:2014nza}. Those measurements aggregate semi-inclusive $A(e,e'p)$ events with two correlated nucleons and a proton momentum in the range $1.5 < p < 3$~fm$^{-1}$.  
    }
    \label{fig:paircomposition}
\end{figure}

Of high relevance for the study of the isospin composition of SRC is that the momentum distribution of Eq.~(\ref{eq:NMD}) can be written as a sum of four terms  
\begin{equation} \label{eq:paircontributiontoNMD}
 n^{[1]}(p) \equiv \underbrace{n^{[1]}_{\text{pp}}(p) \; + \; n^{[1]}_{\text{pn}}(p)}_{n^{[1]}_{\text{p}}(p) \; (\text{proton part})} \; +  \; \underbrace{n^{[1]}_{\text{nn}}(p) \;  + \; n^{[1]}_{\text{np}}(p)}
 _{n^{[1]}_{\text{n}}(p) \; (\text{neutron part})}
 \; .
\end{equation}
Figure~\ref{fig:schememomendis} illustrates that for the nucleons NN$^{\prime}$ an isospin projection on the four combinations of pairs is possible. More details about the derivations are in Ref.~\cite{Vanhalst:2014cqa}.
The $n^{[1]}_{\text{pn}}$ and $n^{[1]}_{\text{np}}$ are both proton-neutron contributions. The difference is that the first type has a leading proton and the second type a leading neutron. The leading nucleon is the nucleon that is tagged at the positions $\vec{r}_1$ and $\vec{r}_1^{\; \prime}$ in Fig.~\ref{fig:schememomendis}. The $n^{[1]}_{\text{p}}(p)$ and $n^{[1]}_{\text{n}}(p)$ express how much the protons and neutrons contribute to the single-nucleon momentum distribution at a momentum $p$.

The relative contribution of $\text{NN}^{\prime}$ pairs to ${n^{[1]}}$ at a given nucleon momentum $p$ is determined by  
\begin{equation}
r_{\text{NN}^{\prime}} (p) \equiv
\frac{n^{[1]}_{\text{NN}^{\prime}}(p)}{n^{[1]}(p)} \; .
\end{equation}
 The LCA contributions to $r_{\text{NN}^{\prime}} (p)$ are schematically shown in diagrams (b)-(e) of Fig.~\ref{fig:schememomendis}.
 
\begin{figure}[ht]
\centering
\includegraphics[viewport=42 28 538 413, clip, width=0.70\columnwidth]{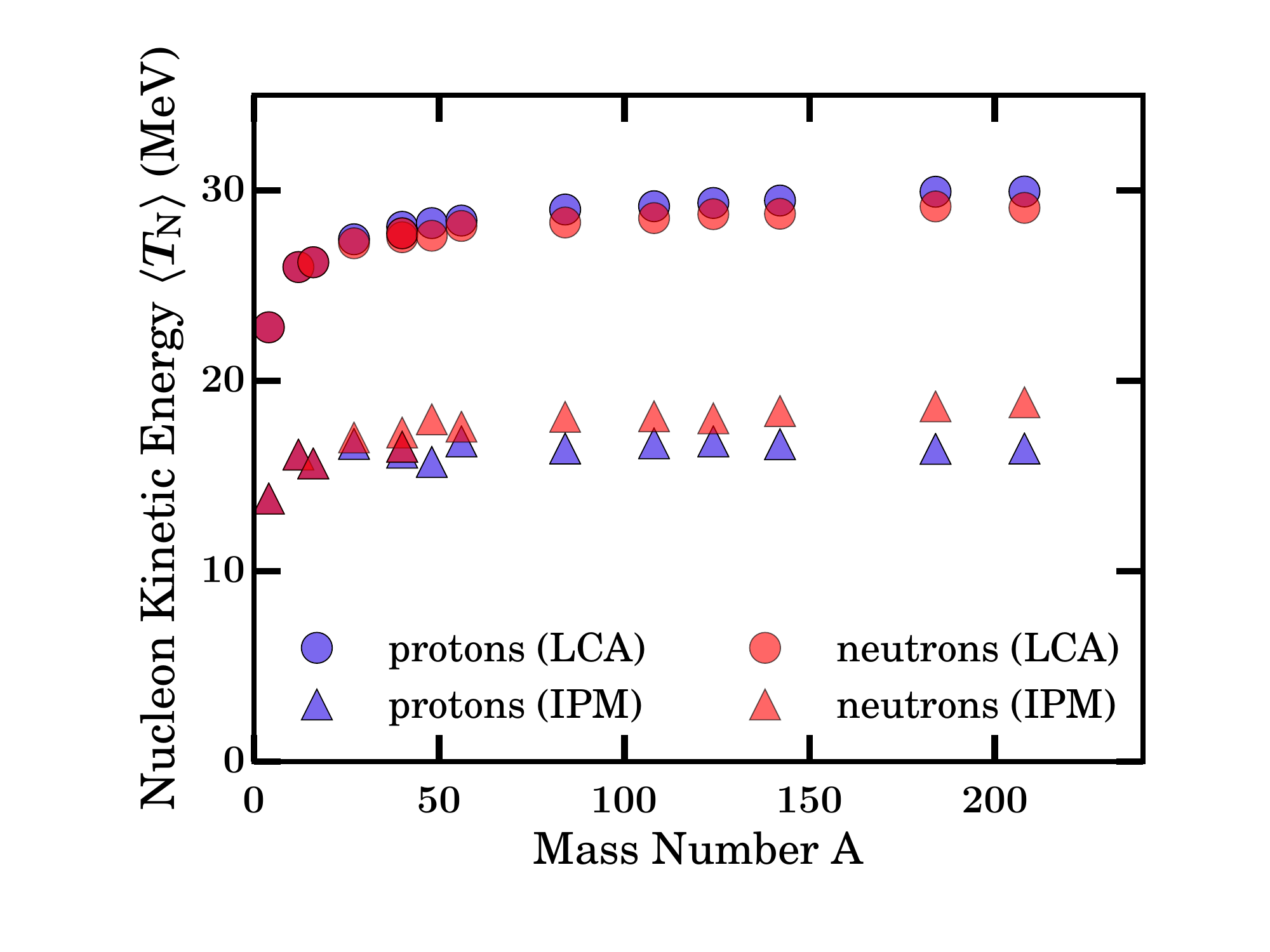}
\includegraphics[viewport=42 28 538 413, clip, width=0.70\columnwidth]{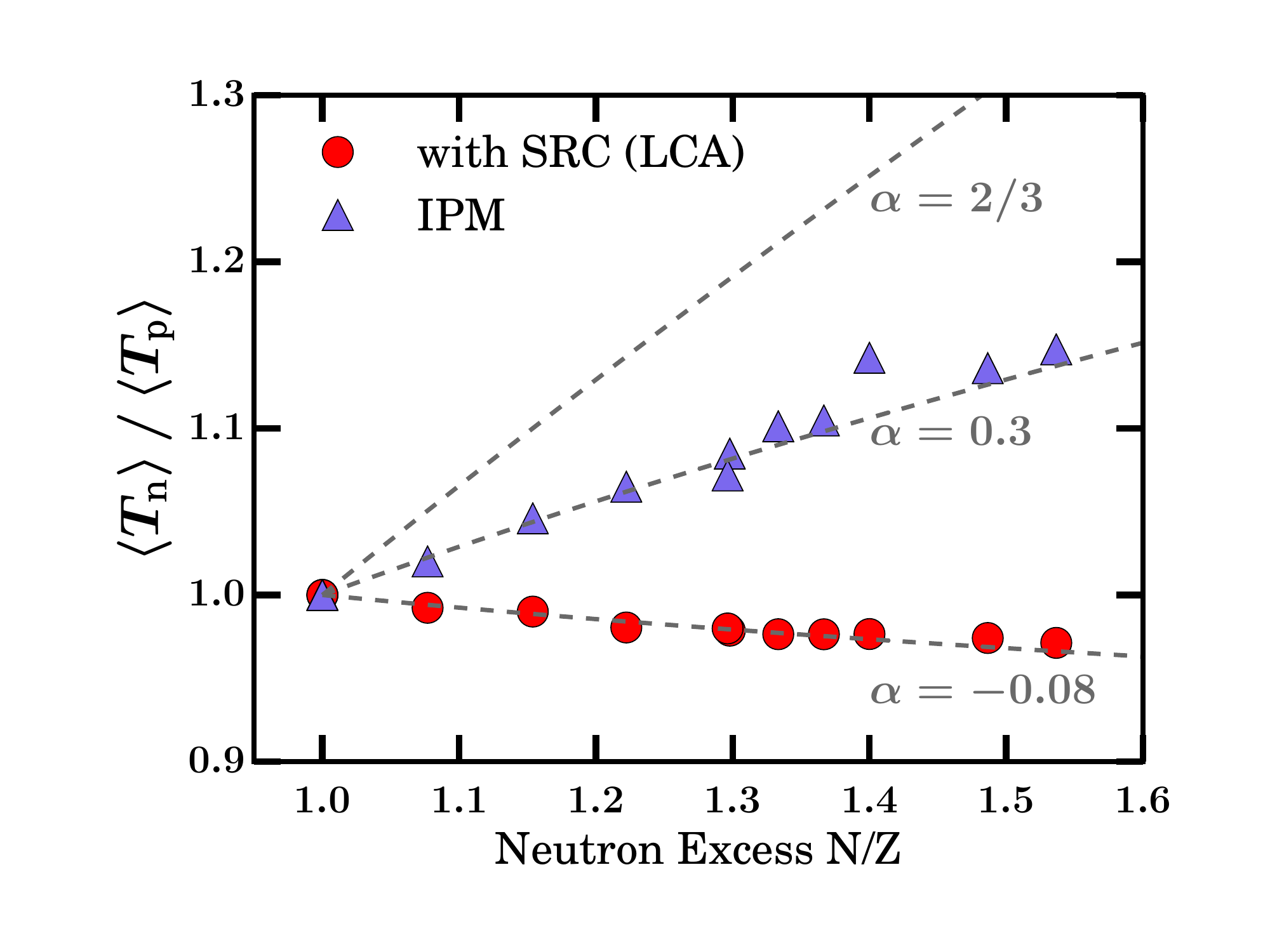}
\caption{Upper panel: The expectation values for the non-relativistic kinetic energy for protons and neutrons as computed from the second moment of the momentum distributions. Results are displayed as a function of the nuclear mass number  for the  nuclei contained in Fig.~\ref{figmomentumdis}. Both the IPM and the LCA (``hard'' $g_c$) results are shown. Lower panel: The ratio of the proton to neutron non-relativistic kinetic energy as a function of the neutron excess $\frac {N} {Z}$. The dashed lines are to guide the eye and represent the function $\left( \frac{N}{Z} \right) ^{\alpha}$ for three values of $\alpha$. } 
\label{fig:ekinresults}
\end{figure}

Figure~\ref{fig:paircomposition} displays the LCA predictions for the nucleon momentum dependence of the $\left( r_{\text{pp}}, r_{\text{nn}}, r_{\text{np}}+r_{\text{pn}} \right)$ for the four target nuclei that have been the subject of the investigations of the CLAS collaboration at Jefferson Laboratory~\cite{Hen:2014nza, Duer2018, Colle:2015ena}.  In the IPM regime (loosely defined as $p<p_F$), the pair composition of the momentum distribution displays rather little momentum dependence and the values for the $r_{\text{NN}^{\prime}}$ are compatible with the naive IPM expectations  
\begin{equation}
r_{\text{pp}}=\frac{Z(Z-1)}{A(A-1)}, r_{\text{nn}}=\frac{N(N-1)}{A(A-1)}, 
r_{\text{pn}}=r_{\text{np}}=\frac{NZ}{A(A-1)} \; .
\end{equation}
In the fat tail, on the other hand, the pair composition of the ${n^{[1]}(p)}$ has a characteristic momentum dependence across all nuclei. At the low momentum side of the fat tail ($2.0 \lesssim p \lesssim 2.75~\text{fm}^{-1}$, shaded area in Fig.~\ref{fig:paircomposition}), pn pairs systematically account for 80-90\% of ${n^{[1]}(p)}$. In that momentum range, pp and nn pairs each account for about 5-10\%.  This is reminiscent of the fact that tensor correlations dominate for $1.5 \lesssim
p \lesssim 3$~fm$^{-1}$. This means that np-SRC create the majority of the high-momentum protons and neutrons in nuclei. The effect of the isospin-blind central correlations becomes substantial at $p \gtrsim  3.5$~fm$^{-1}$ and this is reflected in the corresponding  ${r_{\text{NN}^{\prime}} (p)}$~\cite{Korover:2014dma, Vanhalst:2014cqa}.  
For the ${r_{\text{NN}^{\prime}} (p)}$ the theoretical uncertainties related to the choice of the $g_c$ are at most $5$\%.
The above indicates that the pp and pn SRC pair fractions are momentum dependent. As the tail parts of the momentum distributions are exponentially decreasing functions, semi-exclusive $A(e,e'pN)$ measurements in $p>p_F$ kinematics mainly probe the low $p$ part of the SRC region. In those regions we find that $r_{\text{np}} + r_{\text{pn}} \approx 0.9$ and $r_{\text{nn}} + r_{\text{pp}} \approx 0.1$. Accordingly, the 
LCA method predicts that $\approx$90\% of correlated pairs is pn, and $\approx$5\% is pp. Moreover, there are only smooth variations of those numbers across the nuclear mass table and this prediction is  in line with the experimental results reported in Ref.~\cite{Hen:2014nza}. 

\begin{figure}[ht]
\centering
\includegraphics[viewport=45 75 530 413, clip, width=0.70\columnwidth]{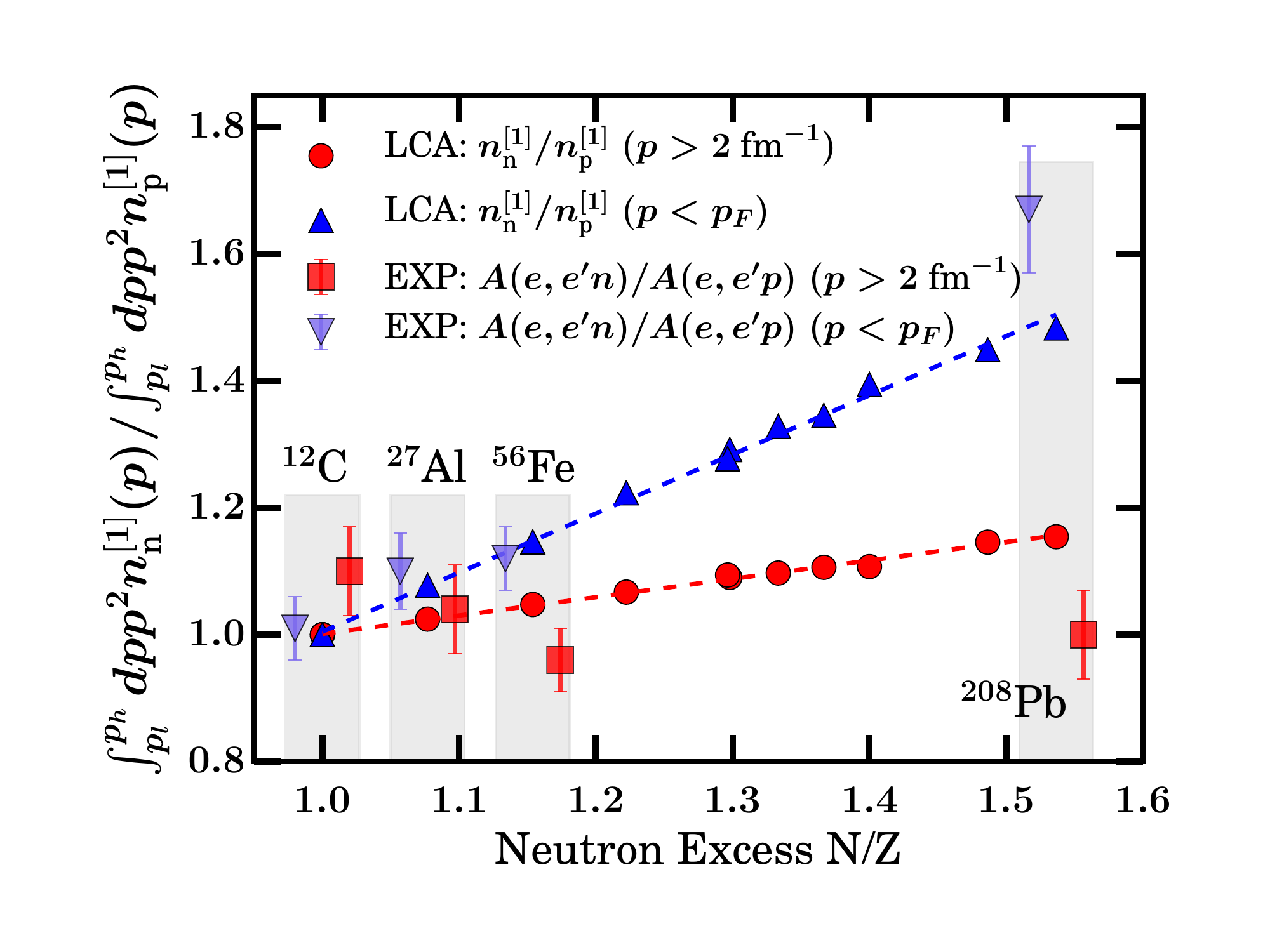}
\includegraphics[viewport=45 28 530 413, clip, width=0.70\columnwidth]{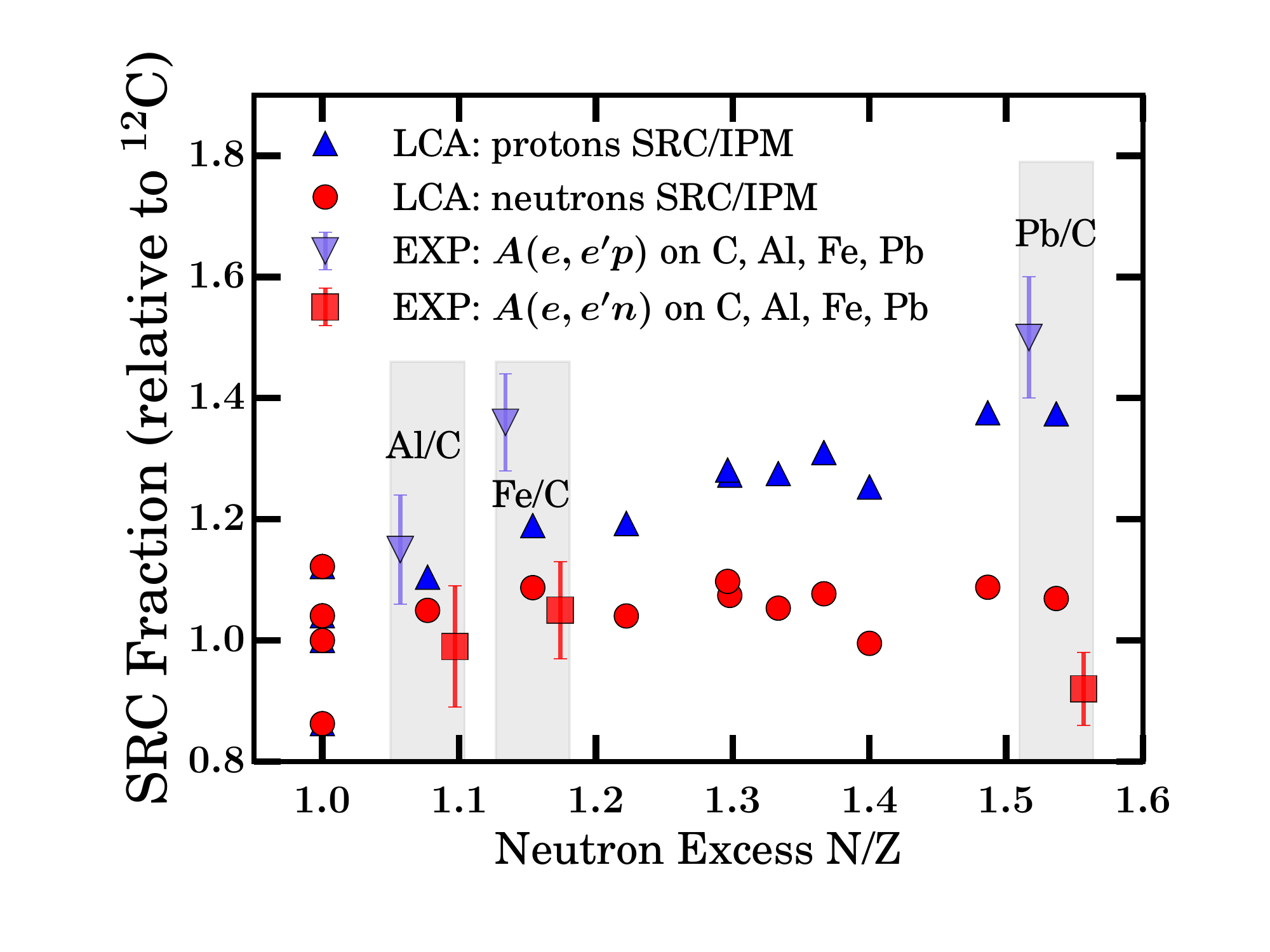}
\caption{Upper panel: The ratio of the weight of the neutrons to the one of the protons  in the momentum distributions in the IPM ($p<p_F$) and SRC ($p>2~\text{fm}^{-1}$) regimes. The LCA (``hard'' $g_c$) results for all 14 nuclei contained in Fig.~\ref{figmomentumdis} are displayed as a function of the neutron excess. The dashed lines are a linear regression to the computed values. The data (slightly displaced horizontally for ease of viewing) are from Ref.~\cite{Duer2018} and  include the target nuclei $^{12}$C, $^{27}$Al, $^{56}$Fe, $^{208}$Pb (shaded boxes). Lower panel: the SRC fraction relative to $^{12}$C as defined in Eq.~(\ref{eq:superratio}) as a function of the  neutron excess $\frac{N}{Z}$.} 
\label{fig:SRCfractions}
\end{figure}

%
%
The second moment of the $n^{[1]}(p)$ is connected with the expectation value of the non-relativistic kinetic energy. The average proton kinetic energy can be readily computed from the pair contributions to $n^{[1]}(p)$ defined in Eq.~(\ref{eq:paircontributiontoNMD})  
\begin{equation}
\langle T_{\textrm{p}}\rangle = \frac {1}{2 m_p} 
 \frac{\int _{0} ^ {\Lambda} dp \; p ^ 4 \;  \left[ n^{[1]}_{\text{pp}}(p) \; + n^{[1]}_{\text{pn}}(p) \right] }{\int _{0} ^ {\Lambda} dp \;  p^2 \;  \left[ n^{[1]}_{\text{pp}}(p) \; + \; n^{[1]}_{\text{pn}}(p) \right]}
  \; ,   
\end{equation}
where $m_p$ is the proton mass. A similar expression holds for $\langle T_{\textrm{n}} \rangle$.  $\Lambda$ is a ultraviolet momentum cutoff for which we use a value lower than the nucleon mass  $(\Lambda=4$~fm$^{-1})$. The CDFs of Fig.~\ref{figcdf} indicate that at $\Lambda = 4$~fm$^{-1}$ almost all probability is exhausted. The results for $\langle T_\textrm{p} \rangle$ and  $\langle T_\textrm{n} \rangle$ and their ratio  are contained in Fig.~\ref{fig:ekinresults} for all 14 nuclei of Fig.~\ref{figmomentumdis}. We have computed the  $\langle T_\textrm{N} \rangle$ in the IPM and the LCA. We find that across the nuclear mass table the SRC accounts for about 40\% of the $\langle T_\textrm{N} \rangle$. With respect to the difference between $\langle T_{\textrm{p}} \rangle$ and $\langle T_{\textrm{n}} \rangle$ we find that the IPM results are fully in line with the expectations: the majority species has the largest value. After including the SRC, however, the situation is reversed with the minority species (protons) having the largest kinetic energy. In other words, per nucleon the minority species (protons) are more short-range correlated than the majority species and the SRC induce inversion of kinetic energy sharing.
The cluster variational Monte Carlo calculations for $^{40}$Ca of Ref.~\cite{Lonardoni:2017egu} are available for two choices of the inter-nucleon interaction and result in $\langle T_\textrm{N} \rangle = 30.34 \pm 1.60$~MeV (AV18+UIX) and $\langle T_\textrm{N} \rangle = 32.54 \pm 1.94$~MeV (AV18). Those calculations integrate the momentum distribution up to $p=10~\text{fm}^{-1}$ and 
fully include the effect of long-range correlations. For $^{40}$Ca, after accounting for SRC in LCA we find $\langle T_{\textrm{p}} \rangle=\langle T_{\textrm{n}} \rangle =27.75$~MeV with the ``hard'' $g_c$ and $\langle T_{\textrm{p}} \rangle=\langle T_{\textrm{n}} \rangle =26.67$~MeV with the ``soft'' $g_c$. For all quantities displayed in Figs.~\ref{fig:ekinresults} and~\ref{fig:SRCfractions} the theoretical uncertainties connected with the choice of $g_c$ are at the percent level which corresponds to the size of the symbols used. This is another indication of the predominant role of the tensor correlations when evaluating quantities that aggregate over nucleon momentum ranges.    

The results for the ratio $\langle T_{\textrm{n}} \rangle / \langle T_{\textrm{p}}\rangle$ in the lower panel of Fig.~\ref{fig:ekinresults}, indicate that the weight of the minority component in the fat-tailed (SRC) part of $n^{[1]}(p)$ strongly depends on the neutron-to-proton ratio $N/Z$ and is heavily affected by the short-distance dynamics. Indeed, in a naive Fermi-gas model one predicts that $\langle T_{\textrm{n}} \rangle / \langle T_{\textrm{p}} \rangle \sim \left( \frac{N}{Z} \right)^{\alpha} $, with $\alpha = \frac{2}{3}$. The IPM results of Fig.~\ref{fig:ekinresults} give rise to a positive $\alpha$ that is considerably smaller ($\alpha \approx 0.3$). After correcting for SRC, one finds values that are negative ($\alpha \approx -0.08$). This is a very nice example of a physically relevant quantity that receives huge corrections from SRC~\cite{PhysRevC.89.034305}. In the forthcoming, we will discuss how one can obtain more detailed information about this phenomenon that is essentially related to the relative weight  of the protons and the neutrons in the momentum distribution in the IPM and the SRC region. Experimental information about this phenomenon could recently be obtained~\cite{Duer2018} by comparing $A(e,e^{\prime}p)$ en $A(e,e^{\prime}n)$ results at low and high missing momenta for the four target nuclei $ ^{12}$C, $^{27}$Al, $^{56}$Fe, $^{208}$Pb. These measurements provided pivotal information about the neutron-to-proton asymmetry dependence of the nuclear SRC. We now discuss in how far the LCA results for the $\frac{N}{Z}$ dependence of the SRC comply with the trends observed in the data. In our discussions we include the four target nuclei that were subject of the experimental investigations, as well as the 10 other nuclei for which the momentum distribution has been obtained (see Fig.~\ref{figmomentumdis}). We investigate in how far the LCA can reproduce the measured $\frac{N}{Z}$ trends (4 nuclei) and check whether the other 10 nuclei obey the inferred systematic properties by essentially filling the gap between $\frac{N}{Z}=\frac{30}{26}$ (Fe) and $\frac{N}{Z}=\frac{126}{82}$ (Pb).    


The weight of the protons and the neutrons in the momentum distribution in a certain momentum range $[p_{l}, p_{h}]$  can be computed from 
\begin{equation}
\int _{p_{l}}^{p_{h}} dp \; p^{2} \; n^{[1]}_{\text{p}} (p) \hspace{0.05\columnwidth} \text{and,} \; 
\int _{p_{l}}^{p_{h}} dp \; p^{2} \; n^{[1]}_{\text{n}} (p) \; .
\end{equation}
   
The relative weight of the protons and the neutrons in the momentum distribution can be ``measured'' by evaluating the ratio of the electroinduced neutron-to-proton knockout cross sections in a detector that approaches a ``$4\pi$" topology (so that a very large fraction of the phase space gets probed). This amounts to evaluating a cross-section ratio of the form 
\begin{equation}
\frac {\sigma_{ep}}{\sigma_{en}}
\frac 
{ \sigma _{A}(e,e^{\prime}n)}
 {  \sigma _{A}(e,e^{\prime}p)  } \sim 
 \frac{\text{Probability to find a neutron in A in~} p \text{~range}}
      {\text{Probability to find a proton in A in~} p \text{~range}}
 \; . 
 \label{eq:neutrontoproton}
 \end{equation}
 Here,  $\sigma_{ep}$ ($\sigma_{en}$) denotes the off-shell electron-proton (electron-neutron) cross section and $\sigma_A (e,e^{\prime}N)$ is the cross section for $N$ knockout aggregated over a certain initial-nucleon momentum range. 
The neutron-to-proton ratio of Eq.~(\ref{eq:neutrontoproton}) has been evaluated at ``low''  and ``high'' missing momenta by the CLAS Collaboration at Jefferson Lab \cite{Duer2018} and the data are included in Fig.~\ref{fig:SRCfractions}. 

Assuming that final-state interactions equally affect neutron and proton knockout, the theoretical counterpart of the ratio (\ref{eq:neutrontoproton}) reads
\begin{equation} \label{eq:thneutrontoproton}
\frac
{\int_{p_{l}}^{p_{h}} dp p^2 n^{[1]}_{\text{n}}(p)}
{\int_{p_{l}}^{p_{h}} dp p^2 n^{[1]}_{\text{p}}(p)} \; .
\end{equation}
Following the experimental cuts used for the results of Ref.~\cite{Duer2018}, the ``IPM'' region is defined through  $p_{l}=0 \le p \le p_{h}=p_F=1.25~$fm$^{-1}$ and the ``SRC'' region through  $p_{l}=2.0 \le p \le p_{h}=5.0~$fm$^{-1}$. 

In Fig.~\ref{fig:SRCfractions} we display the computed asymmetry dependence of the relative weight of neutrons and protons in the $n^{[1]}(p)$ in the IPM and SRC regimes. 
Obviously, the relative weight of the protons and neutrons is very different in ``IPM'' and ``SRC'' regions of the nuclear momentum distributions. Indeed, from Fig.~\ref{fig:SRCfractions} it is clear that the ratios of Eq.~(\ref{eq:thneutrontoproton}) have a linear dependence on the neutron-to-proton excess that is different for the IPM and SRC regions. 
In the IPM regime, corresponding with the bulk part of the momentum distributions,  the computed ratio follows the naive expectation $\frac{N}{Z}$. A linear regression to the computed IPM ratios results in $0.93 \frac{N}{Z}+0.07$. As outlined earlier on, in the fat tail of the momentum distribution, the protons are punching above their weight which results in a ratio that has a very soft dependence on $\frac{N}{Z}$ (a fit to the computed numbers in the SRC regime leads to $0.29 \frac{N}{Z}+0.71$).  
For the most asymmetric nucleus $^{208}$Pb considered here, the neutron-to-proton weight in the IPM regime is compatible with $\frac{126}{82}=1.54$. In the SRC regime, on the other hand, the relative weight is more like $\frac{126}{110}$ clearly pointing to the fact that the relative weight of the minority component is significantly larger than naive expectations. In the SRC regime, the protons in Pb are punching approximately 35\% above their weight.

The predictions for the ratios of  Eq.~(\ref{eq:thneutrontoproton}) in the IPM and SRC regimes are in line with the trends as observed in the data. Our results between $\frac{N}{Z}=1.15$ (Fe) and $\frac{N}{Z}=1.54$ (Pb) provide additional evidence for the linear trend with $\frac{N}{Z}$. The data for Fe and Pb seem to indicate that in the SRC regime the protons and neutrons account for an almost equal share of the momentum distribution. Our predictions reproduce the trend of the minority species systematically gaining a lot of weight relative to the majority species with increasing neutron-to-proton ratio, but the predicted tendency is somewhat weaker than the data seem to suggest.


A measure that can be anticipated to reveal physics about the weight of the protons and neutrons in the bulk $(p \lesssim p_F)$ and tail $(p \gtrsim 2~\text{fm}^{-1})$ parts of the momentum distribution are the so-called proton and neutron SRC fractions $\mathcal{R}^{\text{SRC/IPM}}_{\text{p}}$ and $\mathcal{R}^{\text{SRC/IPM}}_{\text{n}}$ for nucleus $A$
\begin{equation}
 \mathcal{R}^{\text{SRC/IPM}}_{\text{N}} (A) \equiv
 \frac
 {\int_{2~\text{fm}^{-1}} ^ {5~\text{fm}^{-1}} dp p^2 n_{\text{N}}^{[1]}(p)}
 {\int_{0}^{p_F} dp p^2 n_{\text{N}}^{[1]}(p)}
 \; \; \;  (\text{N}\equiv\text{p,n})
\; .
\label{eq:tailtobulforprotonandneutron}
\end{equation}
The above results suggest a different behaviour for protons and neutrons as one changes $\frac{N}{Z}$. A theory-experiment comparison assessing the quantities (\ref{eq:tailtobulforprotonandneutron}) is feasible by evaluating the super-ratio of $A$=Al, Fe, Pb relative to C (that serves as the benchmark for an N=Z nucleus)
\begin{equation}
\text{EXP:} \; \;  
\frac
{\frac{\left. \sigma _{A}(e,e^{\prime}N) \right|_{p>2~\text{fm}^{-1}}}
     {\left. \sigma _{A} (e,e^{\prime}N) \right|_{p<p_F}}}
{\frac{\left. \sigma_{^{12}\text{C}}(e,e^{\prime}N) \right|_{p>2~\text{fm}^{-1}}}
     {\left. \sigma_{^{12}\text{C}}(e,e^{\prime}N) \right|_{p<p_F}}}
\hspace{0.05\columnwidth} \text{TH:} \; \; 
 \frac{\mathcal{R}^{\text{SRC/IPM}}_{\text{N}} (A)}
      {\mathcal{R}^{\text{SRC/IPM}}_{\text{N}} (^{12}\text{C})}
\; \; (\text{N={p},{n}}) \; .
\label{eq:superratio}
\end{equation}
Results are displayed in the bottom panel of Fig.~\ref{fig:SRCfractions}. The LCA predictions for the $\frac{N}{Z}$ dependence of the SRC fraction for protons and neutrons are clearly different and seem to follow the trends extracted from the $(e,e^{\prime}N)$ measurements.

\section{Summary}
\label{sec:conclusions}
We have presented a systematic study of specific features of SRC. Our study includes 14 nuclei, including 10 asymmetric ones and adopts the low-order correlation operator approximation (LCA). The LCA starts from the idea that the majority of SRC corrections to the IPM are of the 2N type.  We have presented results  on (i) the single-nucleon momentum distributions for all those nuclei; (ii) the nucleon momentum and nuclear mass dependence of the isospin composition of SRC; (iii) the proton-to-neutron ratio dependence of SRC; (iv) the nuclear mass and proton-to-neutron dependence of the non-relativistic kinetic energy for protons and neutrons. 

It is shown that the LCA predictions for the fat tail of momentum distributions are in line with those of ab-initio approaches that are available for $A \leq 40$. An attractive feature is that the LCA  is suitable for systematic SRC studies, for example of its isospin and proton-to-neutron dependence. We have included a comparison between the LCA predictions of those trends and data stemming from SRC searches with electroinduced proton and neutron knockout. 
We have shown that nuclear SRC can be captured by universal and robust principles and that the LCA is an efficient way of computing the SRC contributions to the single-nucleon momentum distributions. 

In line with the results of nuclear-matter calculations~\cite{Frick:2004th,PhysRevC.79.064308} we find that there are sizable isospin and neutron-to-proton asymmetries in the SRC. A proton embedded in a neutron-rich nucleus is subject to increasingly stronger SRC corrections, which amplifies its weight in the fat tail by a considerable amount. We consider our model as a zeroth order model suitable for systematic SRC studies over large ranges of nuclei and proton-to-neutron ratios. We find a fair consistency between the predicted trends and the recent experimental $A(e,e^{\prime}pN)$ results that can be connected to the nuclear mass and isospin dependence of SRC. Our study provides further evidence for the fact that SRC induced spatio-temporal fluctuations in nuclei are measurable, are significant and are quantifiable.

\section*{Acknowledgements}
The computational resources (Stevin Supercomputer Infrastructure) and services used in this work were provided by the VSC (Flemish Supercomputer Center), funded by Ghent University, FWO and the Flemish Government – department EWI. We wish to thank O.~Hen, E.~Piasetzky, M.~Sargsian and N.~Jachowicz for helpful discussions.

\section*{References}

\end{document}